\documentclass[fleqn,11pt,DIV=12]{scrartcl}
\usepackage[bitstream-charter]{mathdesign}
\usepackage[ansinew]{inputenc}
\usepackage[T1]{fontenc}
\usepackage[ngerman,english]{babel}
\usepackage{enumitem}
\usepackage{booktabs}
\usepackage{amsmath}
\usepackage[margin=10pt,font=small,]{caption}
\usepackage{subcaption}
\usepackage{graphicx}
\usepackage{microtype}
\usepackage{upgreek}
\usepackage{cite}
\usepackage{slashed}
\usepackage{xcolor}

\usepackage[
    pdftitle={AVL},
    pdfauthor={Buchheim, Hilger, Kaempfer}
    bookmarks=true,
    pdfborder={0 0 0},
    bookmarksopen=true,
    unicode,
    hyperfootnotes=true,
    colorlinks=false,
    pagebackref=false,
    citecolor=blue,
    linkcolor=blue,
    urlcolor=blue,
    breaklinks,
    pdfpagelabels,
    ]{hyperref}

\usepackage[
	toc,
	acronym,
	nomain,
	style=long,
	nonumberlist,
	nowarn,
	nomain,
	]{glossaries}
\makeglossaries
\renewcommand*{\CustomAcronymFields}{%
  name={\the\glsshorttok},
  description={\the\glslongtok},
  first={\noexpand\the\glslongtok\space(\the\glsshorttok)},%
  firstplural={\noexpand\the\glslongtok\noexpand\acrpluralsuffix\space(\the\glsshorttok\noexpand\acrpluralsuffix)},%
  text={\the\glsshorttok},%
  plural={\the\glsshorttok\noexpand\acrpluralsuffix}%
}
\SetCustomStyle
\glsdisablehyper
\newacronym{OPE}{OPE}{operator product expansion}
\newacronym{EV}{EV}{expectation value}
\newacronym{QSR}{QSR}{QCD sum rule}
\newacronym{VEV}{VEV}{vacuum expectation value}
\newacronym{DCSB}{D$\upchi$SB}{dynamical chiral symmetry breaking}
\newacronym{HQE}{HQE}{heavy-quark expansion}

\begin{document}

\title{Algebraic vacuum limits of QCD condensates from in-medium projections of Lorentz tensors}

\author{T. Buchheim, B. K\"{a}mpfer\\[2mm]
				\small{
				Helmholtz-Zentrum Dresden-Rossendorf, PF 510119, D-01314 Dresden, Germany
				}\\[-1mm]
				\small{Technische Universit\"{a}t Dresden, Institut f\"{u}r Theoretische Physik, D-01062 Dresden, Germany
				}
				}

\date{T. Hilger\\[2mm]
			\small{
			University of Graz, Institute of Physics, NAWI Graz, A-8010 Graz, Austria\\[5mm]
			}
			}

\publishers{\small{\today}}

\maketitle

\begin{abstract}
Utilizing the in-medium Lorentz decomposition of operators generating QCD condensates we derive general constraints among the latter ones by the requirement of a smooth transition from medium to vacuum.
In this way we relate the vacuum limits of heretofore unrelated condensates and provide consistency checks for the vacuum saturation hypothesis and the heavy quark mass expansion.
The results are general and depend only on the rank and symmetry of the Lorentz tensors to be decomposed.
The derived prescription enables to uniquely and directly identify operator product expansion contributions which are algebraically specific for in-medium situations and in particular useful for operators with a higher rank, i.\,e.\ larger than three.
Four-quark condensates in mass dimension six are exemplified.
\end{abstract}

\section{Introduction}

The planned forthcoming experiments at NICA \cite{nica}, FAIR \cite{fair} and J-PARC \cite{j-parc} operate in an energy regime, where charm production sets in, either in proton  or anti-proton nucleus or heavy-ion collisions.
Thus, the heavy quark mass regime becomes accessible in a strongly interacting medium.
Such investigations continue previous studies of in-medium modifications of hadrons in the strangeness sector \cite{Friman:2011zz,Hartnack:2011cn,Hayano:2008vn,Leupold:2009kz} and shift the focus onto the charm sector.
On the theory side, this requires profound evaluations incorporating both, finite densities/temperatures and heavy-quark physics.
\glspl{QSR} have turned out to be a suitable tool in this respect \cite{Shifman:1978bx,Reinders:1984sr,Narison:2007spa,Shuryak:1984nq,Colangelo:2000dp,Furnstahl:1992pi}.
That method provides testable predictions for spectral properties of mesons and baryons in vacuum (see
\cite{Gubler:2014pta,Chen:2015kpa,Jiang:2015paa,Wang:2014gwa,Wang:2015mxa,Dominguez:2014vca,Dominguez:2014pga,Zhou:2015ywa,Fariborz:2015vsa,Lucha:2014xla,Lucha:2014nba,Lucha:2015xua}
for recent studies) and in a medium 
\cite{Leupold:1998bt,Klingl:1997kf,Thomas:2005dc,Narison:2012xy,Hilger:2008jg,Thomas:2007gx,Hilger:2010cn,Gubler:2015yna,Kim:2015ywa,Kim:2015xna}%
; even exotic states can be dealt with
\cite{Nielsen:2009uh,Navarra:2006nd,Lee:2007gs,Albuquerque:2008up,Narison:2010pd,Dias:2013xfa}%
.
Moreover, \glspl{QSR} also allow for more general discussions w.\,r.\,t.\ \gls{DCSB} and its restoration in the medium \cite{Dominguez:1989bz,Kapusta:1993hq,Dominguez:2003dr,Kwon:2008vq,Kwon:2010fw,Hilger:2011cq,Hohler:2012xd,Hohler:2013eba,Ayala:2014rka,Dominguez:2014fua}.
Therefore, reliable knowledge about the underlying \gls{OPE} and related technicalities is mandatory.

Besides the investigations in \cite{Hilger:2008jg,Buchheim:2014rpa,Zschocke:2011aa,Hilger:2009kn} the in-medium evaluations of open-flavour \glspl{QSR} are restricted to mass-shift-only scenarios of the particle--anti-particle doublets, whereas particle and anti-particle are artificially kept degenerate in their mass, e.\,g.\ \cite{Wang:2011mj,Azizi:2014bba,Sundu:2014yxa}, and to mass-dimension 4 \gls{OPE} evaluations, e.\,g.\ \cite{Hayashigaki:2000es}.
However, mass shift and splitting of and within the particle and anti-particle doublet are coupled and the first approximation is equivalent to a vanishing (Lorentz-)odd \gls{OPE} of the current-current correlator.
Due to the broken $\mathrm{SO}(4)$ symmetry in an ambient strongly interacting medium, the dispersion relation of an open-flavour correlator couples integrated spectral densities of particle and anti-particle non-linearly to Lorentz-odd and -even parts of the \gls{OPE}.
Thus also the Lorentz-odd \gls{OPE} enters the mass shift \cite{Hilger:2009kn,Hilger:2008jg,Hilger:2010zb,Kampfer:2010vk,Hilger:2010zf,Rapp:2011zz}.
Regarding precision predictions and higher densities and/or temperatures, there has, so far, not been any reliable estimate of the error caused by neglecting the (Lorentz-)odd part of the \gls{OPE}.
Similarly, a second approximation, i.\,e.\ the truncation of the \gls{OPE}, must also be questioned w.\,r.\,t.\ accuracy.
As the \gls{OPE} is an asymptotic, i.\,e.\ divergent, expansion, contributions of higher mass dimension are required to judge convergence and, thus, ensure or probe the desired accuracy.

As non-vanishing chirally odd condensates, such as the chiral condensate, necessitate the dynamical breakdown of chiral symmetry, their medium dependence \cite{Hatsuda:1992bv,Jin:1992id,Fiorilla:2012bc,Goda:2013bka} is of paramount importance \cite{Weise:1994bg,Thomas:2006nx}, not only in the scope of \glspl{QSR} \cite{Rapp:1999ej,Rapp:1999us,Harada:2005br}.
However, as the vanishing of a single condensate does not necessarily imply chiral symmetry restoration (cf.\ \cite{Kanazawa:2015kca} for a recent discussion of a counter example), a more comprehensive view must be envisaged.
The impact of higher order condensates, in particular of four-quark condensates, on $\uprho$ meson and nucleon properties is well-known \cite{Leupold:1998bt,Zschocke:2002mn,Thomas:2007gx,Thomas:2007es,Thomas:2006nk}.
In particular chirally odd four-quark condensates turned out to materially contribute to the spectral shape of the $\uprho$ meson \cite{Hilger:2010cn}.
Furthermore, it has been demonstrated that also the spectral difference of heavy-light quark parity-partners are driven by order parameters \cite{Hilger:2011cq,Buchheim:2015xka,Hilger:2012db}.
In order to obtain a comprehensive picture of the landscape of order parameters of \gls{DCSB} and their respective relations to the spectra of hadrons and medium modifications thereof, four-quark condensates must also be investigated in the scope of heavy-light quark mesons.

In general, QCD condensates are \glspl{EV} of operators which are scalars w.\,r.\,t.\ color-, Dirac-, and Lorentz-indices.
They emerge as decomposition coefficients of non-scalar operator products.
Applying the background field method in Fock-Schwinger gauge in order to evaluate an \gls{OPE}, one naturally encounters also non-scalar operators.
These are then projected on scalar operators.
Thereby the decomposition into Lorentz scalars is special, as only there the difference between vacuum and medium in the scope of an \gls{OPE} comes into play.
On the other hand the number of Lorentz indices to be projected and the number of decomposition terms tends to increase with the mass dimension of the operator.
As a consequence, one faces the problem of an increasing number of numerically yet unknown condensates.
To overcome this, condensates may be ignored or neglected, if good reasons could be provided, or the vacuum saturation hypothesis may be employed from the very beginning.
However, for the $D$ meson this naive approach leads to an \gls{OPE} which does not correctly reproduce the vacuum limit and consequently would not lead to a continuous dependence of the spectrum on density and/or temperature \cite{Buchheim:2014rpa,Buchheim:2014eya,Hilger:2012db}.
The reason is that the additional medium contribution contains scalar operators that are also contained in the vacuum \gls{OPE}.
Retaining not the full additional medium contribution prevents a proper cancellation of terms and destroys the clean vanishing of the medium-specific contribution in vacuum.
In this context, it is important to clearly distinguish between the notions of \emph{medium-specific scalar operators or condensates} on one side and, on the other side, the \emph{medium dependence of a condensate}:
A medium-specific scalar operator has a vanishing \gls{VEV}.
It only occurs in the medium.
In contrast, a vacuum-specific condensate has a non-zero \gls{VEV}.
It already occurs in vacuum.
Both have a medium dependence.

Based on these notions we are going to present a comprehensive formalized framework suitable to deal with higher mass-dimension condensates.
The goal is a decomposition scheme for tensor structures which accomplishes the smooth transition of the \gls{OPE} of current-current correlators in a medium to the vacuum.

Our paper is organized as follows.
In Sec.~\ref{sec:framework}, the decomposition of Lorentz tensors in vacuum and medium-specific parts is introduced along with the associated algebraic vacuum limits.
We choose four-quark condensates entering the \gls{OPE} of meson current-current correlators to exemplify this framework in Sec.~\ref{sec:4q}.
The obtained algebraic vacuum limits of these condensates are checked for compatibility with factorization and heavy-quark expansion in sections \ref{sec:fac} and \ref{sec:hqe}, respectively.
In Sec.~\ref{sec:chiral}, comments on chirally odd condensates are provided.
Finally, we summarize this work in Sec.~\ref{sec:summary}, while further technical details can be found in the Appendices \ref{app:equiv} (comparing two approaches to a medium-specific condensate) and \ref{app:furtherdecomp} (completing the decomposition of dimension 6 quark operators).

\section{General framework}
\label{sec:framework}

QCD condensates constitute the decomposition coefficients of ground state \glspl{EV} $\langle\mathrm{vac}| O_\mu |\mathrm{vac}\rangle$ and/or Gibbs' averages $\langle\!\langle O_\mu \rangle\!\rangle$ of local spin-$n$ operators $O_\mu$.
We use the short hand notation $\langle O_\mu \rangle$ for both in this paragraph.
The label $\mu$ is understood here as a short hand notation for $n$ Lorentz indices $O_{\mu_1\ldots\mu_n}$ (irrespective of the number of space-time dimensions).
The corresponding Lorentz indices are projected onto a set of independent Lorentz tensors collected in the projection vector $\vec l_\mu$ which represents the dependencies of the \gls{EV} under consideration.\footnote{
Note that we do not refer to $\vec l_\mu$ as an element of a vector space, but only make use of the vector notation for the sake of simplicity.}
We now assume that the following decomposition holds:
\begin{equation}
	\label{eq:defdecomp}
	\langle O_\mu \rangle = \vec l_\mu \cdot \vec a \: ,
\end{equation}
where \(\cdot\) denotes the scalar product. It follows that the desired decomposition is given by
\begin{subequations}
\begin{equation} \label{eq:defalgvac}
    \vec a = \left(\vec l \circ \vec l\right)^{-1} \vec l_\nu \langle O^\nu \rangle \: ,
\end{equation}
\begin{equation}
	\label{eq:genproj}
	\langle O_\mu \rangle = \mathrm{Tr}\left[\left(\vec l \circ \vec l\right)^{-1} \left(\vec l_\mu \circ \vec l_\nu\right) \right] \langle O^\nu \rangle \: ,
\end{equation}
\end{subequations}
where $\circ$ is the dyadic product, to which the trace refers to and contracted indices are understood if omitted.
The projection matrix $L \equiv \left(\vec l_\alpha \circ \vec l^\alpha\right) \equiv \left(\vec l \circ \vec l\right)$ is symmetric and
$T_{\mu\nu} \equiv \mathrm{Tr}\left[\left(\vec l \circ \vec l\right)^{-1} \left(\vec l_\mu \circ \vec l_\nu\right) \right]$
is the wanted projection tensor, satisfying $T_{\mu\nu}T^\nu{}_\kappa = T_{\mu\kappa}$.
Obviously, a valid set of Lorentz tensors and a valid projection vector is only given if the projection matrix $L$ is invertible.
Therefore, the components of $\vec l_\mu$ must be linearly independent.
As $L$ is symmetric, $L^{-1}$ is likewise.

In general, one writes down all possible Lorentz tensors which can be constructed from the set on which the \gls{EV} depends and performs the projection for each \gls{EV} independently.
However, as can be seen from \eqref{eq:genproj}, the procedure of decomposing an operator into a set of Lorentz scalars only depends on the tensor rank and is, thus, actually independent of the operator.
Once $T_{\mu\nu}$ is known, it is the same for all operators $O_\mu$.
On the other hand, a specific operator $O_\mu$ may occur in different \glspl{OPE}.
This in turn may impose different symmetry conditions on the operator by contraction of its Lorentz indices with symmetric or antisymmetric tensors.
It is therefore advisable to anti-/symmetrize the Lorentz indices of $\vec l_\mu$.

\subsection{Vacuum}

In vacuum we assume that an operator $O_\mu$ decomposes as
\begin{equation}
\label{eq:defvac}
	\langle\mathrm{vac}| O_\mu |\mathrm{vac}\rangle = \langle\mathrm{vac}| O_\mu |\mathrm{vac}\rangle_0 = \vec l^{(0)}_\mu \cdot \vec a_{0} \: ,
\end{equation}
where the metric tensor $g_{\mu_1\mu_2}$ and the Levi-Civita pseudo tensor $\epsilon_{\mu_1\mu_2\mu_3\mu_4}$ are the only independent Lorentz tensors to construct the projection vector $\vec l^{(0)}_\mu$.
From these, $\vec l^{(0)}_\mu$ can be constructed only for even rank tensors.
Hence, the Lorentz-odd \gls{OPE} vanishes.

\paragraph{Examples}
For a second-rank tensor (two Lorentz indices)  $\langle\mathrm{vac}| O_{\mu_1\mu_2} |\mathrm{vac}\rangle$, only the metric tensor contributes and the decomposition tensor is trivially given by
\begin{equation}
	T^0_{\mu_1\mu_2\,\nu_1\nu_2} = \frac14 g_{\mu_1\mu_2} g_{\nu_1\nu_2} \: .
\end{equation}
Clearly, only symmetric tensors can give non-vanishing condensates.
For a fourth-rank tensor (four Lorentz indices) $\langle\mathrm{vac}| O_{\mu_1\mu_2\mu_3\mu_4} |\mathrm{vac}\rangle$, three combinations of metric tensors and the Levi-Civita symbol serve as decomposition basis yielding
\begin{align}
	\label{eq:vac4ind}
	&T^0_{\mu_1\mu_2\mu_3\mu_4\,\nu_1\nu_2\nu_3\nu_4} = 
		\frac{1}{72} 	\left( 	\begin{array}{c}
														\epsilon_{\mu_1\mu_2\mu_3\mu_4} 	\\
														g_{\mu_1\mu_2}g_{\mu_3\mu_4} 		\\
														g_{\mu_1\mu_3}g_{\mu_2\mu_4} 		\\
														g_{\mu_1\mu_4}g_{\mu_2\mu_3}
													\end{array}
									\right)^\mathrm{T}
		\left( \begin{array}{cccc}
						-3	&			&			& 		\\
								&	5		&	-1	& -1	\\
								&	-1	&	5		&	-1	\\
								&	-1	&	-1	& 5
						\end{array}
		\right)
		\left( \begin{array}{c}
							\epsilon_{\nu_1\nu_2\nu_3\nu_4}	\\
							g_{\nu_1\nu_2}g_{\nu_3\nu_4}			\\
							g_{\nu_1\nu_3}g_{\nu_2\nu_4}			\\
							g_{\nu_1\nu_4}g_{\nu_2\nu_3}
						\end{array}
		\right) \: .	
\end{align}
Due to different anti-/symmetries of the decomposition elements, the projection matrix $L_{0}$ is in block-diagonal form\footnote{The projection matrix $L$ can be cast in diagonal form employing an orthogonal set of tensors for decomposition, i.\,e.\ $\left( \vec l_\mu \circ \vec l^\mu \right) = \mathrm{diag} (b_1,b_2,\ldots)$, which can be generated, e.\,g., by the Gram-Schmidt orthogonalization. However, this merely shifts the problem from inversion of $L$ to finding an orthogonal set of tensors, which themselves have no relation to the properties of the operator under consideration unlike tensors which reflect the anti-/symmetries of particular index pairs of the operator.}.
Equivalently, a set of anti-/symmetrized tensors may be employed, which simplifies the projection tensor to
\begin{multline}
	\label{eq:T4indvacsym}
	T^0_{\mu_1\mu_2\mu_3\mu_4\,\nu_1\nu_2\nu_3\nu_4} = 
		\frac{1}{72} 	\left(\! 	\begin{array}{c}
														\epsilon_{\mu_1\mu_2\mu_3\mu_4} 	\\
														g_{\mu_1\mu_3}g_{\mu_2\mu_4} -	g_{\mu_1\mu_4}g_{\mu_2\mu_3}	\\
														g_{\mu_1\mu_3}g_{\mu_2\mu_4} +	g_{\mu_1\mu_4}g_{\mu_2\mu_3} \\
														g_{\mu_1\mu_2}g_{\mu_3\mu_4}
													\end{array}
									\!\right)^\mathrm{T}
		\\ \times
		\left( \begin{array}{cccc}
						-3	&			&			& 		\\
								&	3		&			& 		\\
								&			&	2		&	-1	\\
								&			&	-1	& 5
						\end{array}
		\right)
		\left(\! \begin{array}{c}
							\epsilon_{\nu_1\nu_2\nu_3\nu_4}	\\
							g_{\nu_1\nu_3}g_{\nu_2\nu_4} - g_{\nu_1\nu_4}g_{\nu_2\nu_3}		\\
							g_{\nu_1\nu_3}g_{\nu_2\nu_4} + g_{\nu_1\nu_4}g_{\nu_2\nu_3}		\\
							g_{\nu_1\nu_2}g_{\nu_3\nu_4}
						\end{array}
		\!\right) \: .
\end{multline}

\subsection{Medium}

At non-zero density and/or temperature the medium four-velocity $v_\mu$ provides a new element allowing for a number of additional Lorentz tensors to be projected onto (cf.\ \cite{Buchheim:2014ija} for a list of tensors up to 5 Lorentz indices).
We assume that in medium the following decomposition holds for any operator $O_\mu$:
\begin{equation}
    \label{eq:defplain}
    \langle\!\langle O_\mu \rangle\!\rangle = \langle\!\langle O_\mu \rangle\!\rangle_\rho = \vec l^{(\rho)}_\mu \cdot \vec a_{\rho} \: ,
\end{equation}
where $\rho$ is a generalized medium parameter, i.\,e.\ $\rho=0$ in vacuum, and $\rho>0$ in the medium.
The projection vector $\vec l^{(\rho)}_\mu = ( \vec l^{(\rho_0)}_\mu , \vec l^{(\rho_1)}_\mu )^\mathrm{T}$ contains all tensors which already occur in vacuum, i.\,e.\ $\vec l^{(\rho_0)}_\mu \equiv \vec l^{(0)}_\mu$.
All tensors which incorporate the medium four-velocity $v_\mu$ are collected in $\vec l^{(\rho_1)}_\mu$.
Analogously, $\vec a_{\rho} =  ( \vec a_{\rho_0} , \vec a_{\rho_1} )^\mathrm{T}$, but $\vec a_{\rho_0} \neq \vec a_{0}$ due to $\vec l^{(0)} \circ  \vec l^{(\rho_1)} \neq 0$.

The coefficient vectors are given as
\begin{subequations} \label{eq:plaincoefcoupled}
\begin{align}
    \vec a _{\rho_0}
    & = L_0^{-1} \left( \vec l_\nu^{(0)} \langle\!\langle O^\nu \rangle\!\rangle - L_{0,\rho_1} \vec a_{\rho_1} \right) \: ,
    \\
    \vec a _{\rho_1}
    & = L_{\rho_1}^{-1} \left( \vec l_\nu^{({\rho_1})} \langle\!\langle O^\nu \rangle\!\rangle - L_{\rho_1,0} \vec a_\nu^{\rho_0} \right) 
\end{align}
\end{subequations}
and in disentangled form they read
\begin{subequations}\label{eq:plaincoef}
\begin{align}
\vec a _{\rho_0}
& = \left( L_0 - L_{0,\rho_1} L_{\rho_1}^{-1} L_{\rho_1,0} \right)^{-1} \left( \vec l_\nu^{(0)} - L_{0,\rho_1} L_{\rho_1}^{-1} \vec l_\nu^{(\rho_1)} \right) \langle\!\langle O^\nu \rangle\!\rangle \: ,
\\
\vec a _{\rho_1}
& = \left( L_{\rho_1} - L_{{\rho_1},0} L_{0}^{-1} L_{0,{\rho_1}} \right)^{-1} \left( \vec l_\nu^{({\rho_1})} - L_{\rho_1,0} L_{0}^{-1} \vec l_\nu^{(0)} \right) \langle\!\langle O^\nu \rangle\!\rangle \: ,
\end{align}
\end{subequations}
where the sub-matrices $L_0$, $L_{0,\rho_1}$, $L_{\rho_1,0}$ and $L_{\rho_1}$ of the projection matrix $L_\rho = \left( \vec l^{(\rho)} \circ \vec l^{(\rho)} \right)$ are given explicitly in Eq.~\eqref{eq:totprojmatrix} below.
As can be seen, having pre-evaluated the vacuum decomposition is of little use.
Both decomposition vectors have to be evaluated and will, in general, contain all occurring condensates, irrespective of their vacuum limit.

Naturally, one wishes to reduce evaluational efforts in case the vacuum decomposition is already at hand and only the additional medium-specific contribution has to be determined.
At this stage, however, the full decomposition has to be performed and the vacuum contribution subtracted in order to identify the medium-specific term.
In particular, the construction of a medium-specific projection vector $\vec l^{(1)}_\mu$ which directly gives the additional terms is wanted.

We now define the \emph{algebraic} vacuum decomposition in full analogy to \eqref{eq:defalgvac} and \eqref{eq:defvac} as
\begin{equation}
    \label{eq:genprojalg}
    \langle\!\langle O_\mu \rangle\!\rangle_0 = \vec l^{(0)}_\mu \cdot \vec a_{0}
        \equiv \mathrm{Tr}\left[L_0^{-1} \left(\vec l^{(0)}_\mu \circ \vec l^{(0)}_\nu\right) \right] \langle\!\langle O^\nu \rangle\!\rangle \: ,
\end{equation}
which is algebraically given by the same operators as the decomposition \eqref{eq:defvac}.
However, in general, the \glspl{VEV} are now medium dependent.
Here and in the following, \emph{algebraic} means that the projection vector $\vec l_\mu$ is specified and the projection tensor $T_{\mu\nu}$ is applied to the \gls{EV} irrespective if it is the \gls{VEV} or Gibbs average.
We explicitly separate the algebraic vacuum specific terms
\begin{equation}
\label{eq:defmed}
	\langle\!\langle O_\mu \rangle\!\rangle
        \equiv \langle\!\langle O_\mu \rangle\!\rangle_0 + \langle\!\langle O_\mu \rangle\!\rangle_1
        \equiv \vec l^{(0)}_\mu \cdot \vec a_0 + \vec l^{(1)}_\mu \cdot \vec a_{1} \: ,
\end{equation}
which, together with Eq.\,\eqref{eq:defvac}, \emph{defines} the \emph{algebraic} medium-specific contribution $\langle\!\langle O_\mu \rangle\!\rangle_1$.

Note that \eqref{eq:defmed} is an algebraic separation.
This means, that the vacuum-specific condensates generated by the vacuum-specific decomposition \eqref{eq:defvac} exhibit a density dependence when entering the in-medium decomposition in \eqref{eq:defmed}.
Furthermore, it is important to emphasize that the medium-specific projection vector $\vec l^{(1)}_\mu$ is not only constituted by the medium four-velocity or tensors that contain the latter.
The elements in $\vec l^{(0)}_\mu$ are completely contained in $\vec l^{(1)}_\mu$.
Hence, the medium-specific decomposition contribution and the medium-specific condensates are those contributions to \eqref{eq:defmed} which vanish in vacuum.

\begin{figure}%
\includegraphics[width=0.47\columnwidth]{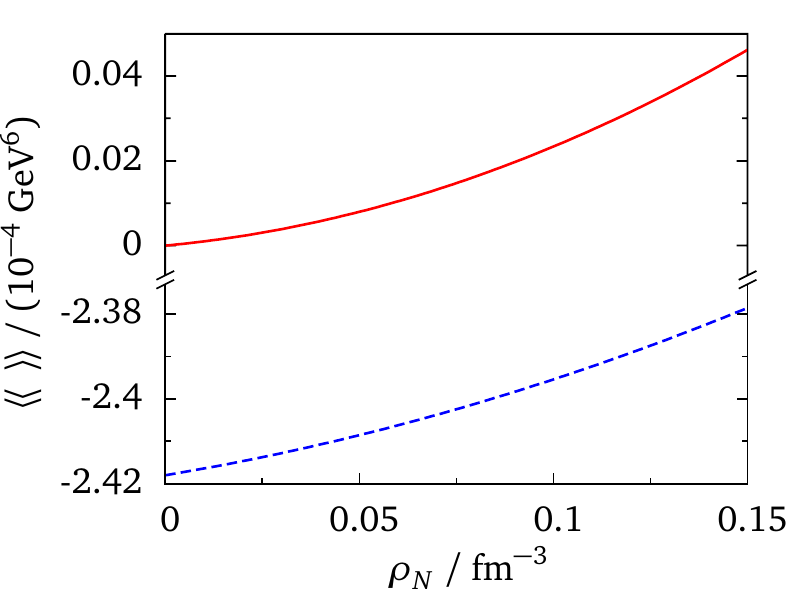}%
\hfill
\includegraphics[width=0.47\columnwidth]{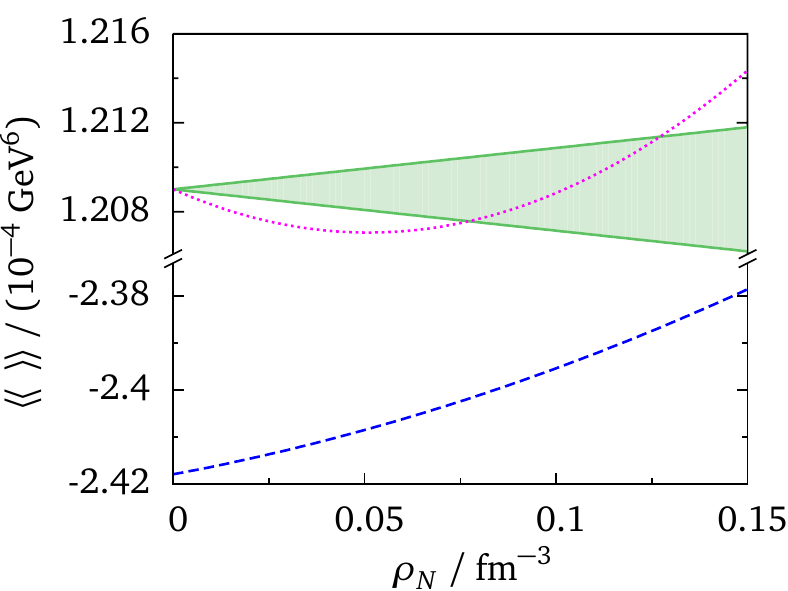}%
\caption{
Left panel: Density dependence of the vacuum-specific condensate $\langle\!\langle g \bar q \gamma_\mu t^A q \sum_f \bar f \gamma^\mu t^A f \rangle\!\rangle$ (blue dashed curve) and the medium-specific condensate $\langle\!\langle g \bar q \gamma_\mu t^A q \sum_f \bar f \gamma^\mu t^A f - 2/v^2 \times \left( \bar q \gamma^\alpha[(vD),G_{\alpha\beta}]v^\beta q + g \bar q \slashed{v} t^A q \sum_f \bar f \slashed{v} t^A f \right) \rangle\!\rangle$ (red solid curve). Right panel: The contributions to the medium-specific condensate, $-2 \langle\!\langle g \bar q \slashed{v} t^A q \sum_f \bar f \slashed{v} t^A f \rangle\!\rangle/v^2$ (magenta dotted curve), $-2 \langle\!\langle\bar q \gamma^\alpha[(vD),G_{\alpha\beta}]v^\beta q\rangle\!\rangle/v^2$ (green band, generated by an assumed linear density dependence $m\,\rho_N$ with $m \in [-2,2]$) and vacuum-specific condensate $\langle\!\langle g \bar q \gamma_\mu t^A q \sum_f \bar f \gamma^\mu t^A f \rangle\!\rangle$ (blue dashed curve, as in left panel), which add up to the red solid curve displayed in the left panel for $m=1$.
}%
\label{fig:fac}%
\end{figure}
To illustrate this important point let us consider an example which is detailed in Sec.~\ref{sec:fac} below.
In Fig.~\ref{fig:fac}, left panel, the vacuum-specific condensate $\langle\!\langle g \bar q \gamma_\mu t^A q \sum_f \bar f \gamma^\mu t^A f \rangle\!\rangle$ is exhibited as a function of the net nucleon density $\rho_N$ by the blue dashed curve.
At $\rho_N=0$ it takes the value $-2.42\times 10^{-4}\;\mathrm{GeV}^6$. According to our definition, this condensate obeys a density dependence, which is deduced here from the squared chiral condensate in nuclear matter.
By definition, the medium-specific condensate $\langle\!\langle g \bar q \gamma_\mu t^A q \sum_f \bar f \gamma^\mu t^A f - \frac{2}{v^2}\left( \bar q \gamma^\alpha[(vD),G_{\alpha\beta}]v^\beta q + g \bar q \slashed{v} t^A q \sum_f \bar f \slashed{v} t^A f \right) \rangle\!\rangle$, depicted as the red solid curve, vanishes at $\rho_N=0$.
Also this condensate obeys a density dependence.
The individual contributions to the medium-specific condensate are the medium four-quark condensate $\langle\!\langle g \bar q \slashed{v} t^A q \sum_f \bar f \slashed{v} t^A f \rangle\!\rangle/v^2$, the quark-gluon condensate $\langle\!\langle\bar q \gamma^\alpha[(vD),G_{\alpha\beta}]v^\beta q\rangle\!\rangle/v^2$ and the already mentioned vacuum four-quark condensate $\langle\!\langle g \bar q \gamma_\mu t^A q \sum_f \bar f \gamma^\mu t^A f \rangle\!\rangle$.
These contributions, depicted as magenta dotted, green solid and blue dashed curves, respectively, in the right panel of Fig.~\ref{fig:fac} add up to the medium-specific condensate shown in the left panel.

This example should illustrate how the vacuum-specific and medium-specific condensates are interwoven in our below presented formalism which has the benefit of displaying a transparent limit $\rho \rightarrow 0$.\\

Using \eqref{eq:plaincoefcoupled} together with \eqref{eq:genprojalg}, equality of the definitions \eqref{eq:defmed} and \eqref{eq:defplain} leads to the following orthogonality requirement for $\vec l^{(0)}_\mu$ and the medium-specific projection vector
\begin{equation}
    \label{eq:defmed2}
    0 = \vec l^{(1)} \circ \vec l^{(0)} \: .
\end{equation}
It can be shown that \eqref{eq:defmed} and \eqref{eq:defmed2} are equivalent definitions (cf.\ Appendix~\ref{app:equiv}).

The according prescription is detailed in the following.
From orthogonality \eqref{eq:defmed2}, the relation $\vec a_0 = L_0^{-1} \left(\vec l^{(0)} \circ \vec l^{(\rho)}\right) \vec a_\rho$ follows, which imposes $\mathrm{dim}(\vec l^{(0)}_\mu)$ constraints on the coefficients $\vec a^{(\rho)}$ and $\vec a^{(0)}$ leaving $\mathrm{dim}(\vec l^{(\rho)}_\mu)$ independent elements.
Substituting the obtained constraints
\begin{equation}
	\label{eq:coeffrel}
	\vec a_{\rho_0} = \vec a_0 - L_0^{-1} L_{0,\rho_1} \vec a_{\rho_1}
\end{equation}
in definition \eqref{eq:defmed} allows (together with Eq.~\eqref{eq:defvac}) to construct a medium-specific projection vector
\begin{equation}
	\label{eq:medprojvec}
	\vec l^{(1)}_\mu = L_{\rho_1,0} L_0^{-1} \vec l^{(0)}_\mu - \vec l^{(\rho_1)}_\mu \: .
\end{equation}
Note that $\vec l^{(1)}_\mu$ is not unique.
A constant factor and symmetry properties may be chosen differently.
Due to orthogonality \eqref{eq:defmed2}, vacuum and medium-specific parts of the tensor decomposition can be evaluated independently of each other and the general decomposition procedure \eqref{eq:genproj} can be applied utilizing the medium-specific projection vector \eqref{eq:medprojvec}.

The medium-specific contribution $\langle\!\langle O_\mu \rangle\!\rangle_1$ of a decomposed Gibbs averaged operator is also referred to as a higher-twist contribution $\langle\!\langle \mathcal{ST} O_\mu \rangle\!\rangle$, where $\mathcal{ST}$ renders the operator symmetric and traceless w.\,r.\,t.\ Lorentz indices \cite{Hatsuda:1992bv,Leupold:1998bt}.
The numerical values of these quantities can be obtained from DIS amplitudes \cite{Gubler:2015uza}. 
In the very same fashion numerical values can be found for medium-specific condensates or at least for linear combinations of the latter.
For medium-specific condensates up to mass dimension 5 the corresponding numerical values and their medium behaviors from DIS amplitudes to leading order in the nucleon density are explicated in \cite{Jin:1992id}.

\paragraph{Examples}
For a second-rank tensor (two Lorentz indices) $\langle\!\langle O_{\mu_1\mu_2} \rangle\!\rangle_1$, only the metric tensor $g_{\mu_1\mu_2}$ and $v_{\mu_1} v_{\mu_2}$ contribute, and the decomposition tensor reads
\begin{align}
	T^1_{\mu_1\mu_2\,\nu_1\nu_2} = \frac{1}{12} \bigg( g_{\mu_1\mu_2} - 4\frac{v_{\mu_1} v_{\mu_2}}{v^2} \bigg) \bigg( g_{\nu_1\nu_2} - 4\frac{v_{\nu_1} v_{\nu_2}}{v^2} \bigg) \: ,
\end{align}
where non-vanishing results are obtained for symmetric Lorentz indices only.
For a fourth-rank tensor (four Lorentz indices) $\langle\!\langle O_{\mu_1\mu_2\mu_3\mu_4} \rangle\!\rangle_1$ fourteen tensors contribute.
Four of them, which incorporate the Levi-Civita symbol, can be evaluated separately (cf.\ Eq.~\eqref{eq:vac4ind}).
Thus, we restrict the consideration to the remaining ten tensors, giving
\begin{multline}
	T^1_{\mu_1\mu_2\mu_3\mu_4\,\nu_1\nu_2\nu_3\nu_4} = \frac{1}{240} \\ \times \mathrm{Tr} \left[ \left( 
	\begin{array}{ccccccc}
		           7 & -\frac{1}{2} & -\frac{1}{2} &            2 & -\frac{1}{2} & -\frac{1}{2} &          -1 \\
		-\frac{1}{2} &            7 & -\frac{1}{2} & -\frac{1}{2} &            2 & -\frac{1}{2} &          -1 \\
		-\frac{1}{2} & -\frac{1}{2} &            7 & -\frac{1}{2} & -\frac{1}{2} &            2 &          -1 \\
		           2 & -\frac{1}{2} & -\frac{1}{2} &            7 & -\frac{1}{2} & -\frac{1}{2} &          -1 \\
		-\frac{1}{2} &            2 & -\frac{1}{2} & -\frac{1}{2} &            7 & -\frac{1}{2} &          -1 \\
		-\frac{1}{2} & -\frac{1}{2} &            2 & -\frac{1}{2} & -\frac{1}{2} &            7 & 				 -1 \\
		          -1 &           -1 &           -1 &           -1 &           -1 &           -1 & \frac{4}{3} 
	\end{array}
  \right) \left( \vec l^{(1)}_{\mu_1\mu_2\mu_3\mu_4} \circ \vec l^{(1)}_{\nu_1\nu_2\nu_3\nu_4} \right) \right]
\end{multline}
with
\begin{align}
	\vec l^{(1)}_{\mu_1\mu_2\mu_3\mu_4} = \bigg( & g_{\mu_1\mu_2}g_{\mu_3\mu_4} - 4g_{\mu_1\mu_2}\frac{v_{\mu_3} v_{\mu_4}}{v^2}, g_{\mu_1\mu_3}g_{\mu_2\mu_4} - 4g_{\mu_1\mu_3}\frac{v_{\mu_2}v_{\mu_4}}{v^2}, \nonumber\\
	& g_{\mu_1\mu_4}g_{\mu_2\mu_3} - 4g_{\mu_1\mu_4}\frac{v_{\mu_2} v_{\mu_3}}{v^2}, g_{\mu_3\mu_4}g_{\mu_1\mu_2} - 4g_{\mu_3\mu_4}\frac{v_{\mu_1} v_{\mu_2}}{v^2}, \nonumber\\
	& g_{\mu_2\mu_4}g_{\mu_1\mu_3} - 4g_{\mu_2\mu_4}\frac{v_{\mu_1} v_{\mu_3}}{v^2}, g_{\mu_2\mu_3}g_{\mu_1\mu_4} - 4g_{\mu_2\mu_3}\frac{v_{\mu_1} v_{\mu_4}}{v^2}, \nonumber
\end{align}
\begin{align}
	\phantom{\vec l^{(1)}_{\mu_1\mu_2\mu_3\mu_4} = \bigg(}
	& g_{\mu_1\mu_2}g_{\mu_3\mu_4} + g_{\mu_1\mu_3}g_{\mu_2\mu_4} + g_{\mu_1\mu_4}g_{\mu_2\mu_3} -  24\frac{v_{\mu_1} v_{\mu_2} v_{\mu_3} v_{\mu_4}}{v^4} \bigg)^\mathrm{T}
\end{align}
and $\vec l^{(1)}_{\nu_1\nu_2\nu_3\nu_4}$ analogously.
Decomposing the components of the projection vector into tensors symmetric and anti-symmetric in the index pairs $\mu_1\mu_2$ and $\mu_3\mu_4$ yields
\begin{multline}
	\label{eq:T4indmedsym}
	T^1_{\mu_1\mu_2\mu_3\mu_4\,\nu_1\nu_2\nu_3\nu_4} = \frac{1}{240} \\ \times \mathrm{Tr} \left[ \left( 
	\begin{array}{ccccccc}
		          10 & 						  & 					   &              & 					   & 						  &             \\
							   &           20 & 					   & 						  &              & 						  &             \\
							   & 						  &           20 &              & 					   &              &             \\
		             & 						  & 					   &            8 & 					-1 & 					-1 &           -2 \\
							   &              & 					   & 					 -1 &            7 &            2 &           -1 \\
							   & 					    &              & 					 -1 &            2 &            7 & 				  -1 \\
		             &              &              &           -2 &           -1 &           -1 & \frac{4}{3} 
	\end{array}
  \right) \left( \vec l^{(1)}_{\{\mu_1\mu_2\}\{\mu_3\mu_4\}} \circ \vec l^{(1)}_{\{\nu_1\nu_2\}\{\nu_3\nu_4\}} \right) \right]
\end{multline}
with
\begin{align}
	\vec l^{(1)}_{\{\mu_1\mu_2\}\{\mu_3\mu_4\}} & = \bigg( l^{(1)}_{[\mu_1\mu_2][\mu_3\mu_4]}, l^{(1)}_{(\mu_1\mu_2)[\mu_3\mu_4]}, l^{(1)}_{[\mu_1\mu_2](\mu_3\mu_4)}, \vec l^{(1)}_{(\mu_1\mu_2)(\mu_3\mu_4)} \bigg)
\end{align}
\begin{subequations}
\begin{align}
	\label{eq:med4indProjVecSym}
	l^{(1)}_{[\mu_1\mu_2][\mu_3\mu_4]} & = g_{\mu_1\mu_3}g_{\mu_2\mu_4} - g_{\mu_1\mu_4}g_{\mu_2\mu_3} - \frac{2}{v^2}\left( g_{\mu_1\mu_3} v_{\mu_2} v_{\mu_4} \right. \nonumber\\
	& \phantom{=} \left. - g_{\mu_1\mu_4} v_{\mu_2} v_{\mu_3} + g_{\mu_2\mu_4} v_{\mu_1} v_{\mu_3} - g_{\mu_2\mu_3} v_{\mu_1} v_{\mu_4} \right) \: , \\
	l^{(1)}_{(\mu_1\mu_2)[\mu_3\mu_4]} & = \frac{1}{v^2}\left( g_{\mu_1\mu_3} v_{\mu_2} v_{\mu_4} - g_{\mu_1\mu_4} v_{\mu_2} v_{\mu_3} - g_{\mu_2\mu_4} v_{\mu_1} v_{\mu_3} + g_{\mu_2\mu_3} v_{\mu_1} v_{\mu_4} \right) \: , \\
	l^{(1)}_{[\mu_1\mu_2](\mu_3\mu_4)} & = \frac{1}{v^2}\left( g_{\mu_1\mu_3} v_{\mu_2} v_{\mu_4} + g_{\mu_1\mu_4} v_{\mu_2} v_{\mu_3} - g_{\mu_2\mu_4} v_{\mu_1} v_{\mu_3} - g_{\mu_2\mu_3} v_{\mu_1} v_{\mu_4} \right) \: , \\
	\vec l^{(1)}_{(\mu_1\mu_2)(\mu_3\mu_4)} & = \bigg( g_{\mu_1\mu_3}g_{\mu_2\mu_4} + g_{\mu_1\mu_4}g_{\mu_2\mu_3} - \frac{2}{v^2}\left( g_{\mu_1\mu_3} v_{\mu_2} v_{\mu_4} \right. \nonumber\\
	& \phantom{= \bigg(} \left. + g_{\mu_1\mu_4} v_{\mu_2} v_{\mu_3} + g_{\mu_2\mu_4} v_{\mu_1} v_{\mu_3} + g_{\mu_2\mu_3} v_{\mu_1} v_{\mu_4} \right), \nonumber\\
	& \qquad\! g_{\mu_1\mu_2}g_{\mu_3\mu_4} - 4g_{\mu_1\mu_2}\frac{v_{\mu_3} v_{\mu_4}}{v^2}, g_{\mu_1\mu_2}g_{\mu_3\mu_4} - 4g_{\mu_3\mu_4}\frac{v_{\mu_1} v_{\mu_2}}{v^2}, \nonumber\\
	& \qquad\! g_{\mu_1\mu_2}g_{\mu_3\mu_4} + g_{\mu_1\mu_3}g_{\mu_2\mu_4} + g_{\mu_1\mu_4}g_{\mu_2\mu_3} -  24\frac{v_{\mu_1} v_{\mu_2} v_{\mu_3} v_{\mu_4}}{v^4}	\bigg)^\mathrm{T}
\end{align}
\end{subequations}
and $\vec l^{(1)}_{\{\nu_1\nu_2\}\{\nu_3\nu_4\}}$ analogously, where the Bach bracket notation is employed to the bracketed index pair $\{\cdots\}$.

Due to the anti-symmetry of the indices of the gluon field strength tensor, the medium-specific gluon condensate can be identified by the decomposition \cite{Zschocke:2011aa}
\begin{multline}
	\langle\!\langle G^A_{\mu_1\mu_2}G^A_{\mu_3\mu_4} \rangle\!\rangle_1 = \frac{1}{24} \langle\!\langle \left( \frac{G^2}{4} - \frac{(vG)^2}{v^2} \right) \rangle\!\rangle \bigg[ g_{\mu_1\mu_3}g_{\mu_2\mu_4} - g_{\mu_1\mu_4}g_{\mu_2\mu_3}\\ - \frac{2}{v^2}\left( g_{\mu_1\mu_3} v_{\mu_2} v_{\mu_4} - g_{\mu_1\mu_4} v_{\mu_2} v_{\mu_3} + g_{\mu_2\mu_4} v_{\mu_1} v_{\mu_3} - g_{\mu_2\mu_3} v_{\mu_1} v_{\mu_4} \right) \bigg]
\end{multline}
which features only the decomposition in Eq.~\eqref{eq:med4indProjVecSym}.

\subsection{Vacuum constraints}

Although the anti-/symmetrized projection vector exhibits a more complicated structure, it considerably simplifies the decomposition of tensors with known symmetries among the Lorenz indices and allows for an unambiguous identification of medium-specific condensates.

In vacuum, Gibbs averaging reduces to the \gls{VEV} of the operators under consideration, \(\langle\!\langle O_\mu \rangle\!\rangle \to \langle \mathrm{vac} | O_\mu | \mathrm{vac} \rangle\).
Hence, in the vacuum limit the non-orthogonal decomposition must satisfy
\begin{equation}
    \vec l^{(0)}_\mu \cdot \vec a_{\rho_0}
    + \vec l^{(\rho_1)}_\mu \cdot \vec a_{\rho_1} \to \vec l^{(0)}_\mu \cdot \vec a_{0} \: ,
\end{equation}
which can be expressed by the following vacuum constraint
\begin{equation}\label{eq:vacconstr_nonort1}
    \binom{\vec l^{(0)}_\mu}{\vec l^{(\rho_1)}_\mu}^\mathrm{T} \left[
        L_\rho^{-1} -
        \left(\begin{array}{cc} L_0^{-1} & 0 \\ 0 & 0 \end{array}\right) \right]
        \binom{\vec l^{(0)}_\nu}{\vec l^{(\rho_1)}_\nu} \langle \mathrm{vac} | O^\nu | \mathrm{vac} \rangle
        =0
        \: .
\end{equation}
As the components of $\vec l^{(\rho)}_\mu$ must be linearly independent, \eqref{eq:vacconstr_nonort1} can only be fulfilled if
\begin{equation}\label{eq:vacconstr_nonort2}
    \left[
        L_\rho^{-1} -
        \left(\begin{array}{cc} L_0^{-1} & 0 \\ 0 & 0 \end{array}\right) \right]
        \binom{\vec l^{(0)}_\nu}{\vec l^{(\rho_1)}_\nu} \langle \mathrm{vac} | O^\nu | \mathrm{vac} \rangle
        =0 \: .
\end{equation}
On the other hand $\vec l_\mu^{(\rho)} \langle \mathrm{vac} | O^\mu | \mathrm{vac} \rangle \neq 0$ must hold, if $\vec l_\mu^{(0)} \langle \mathrm{vac} | O^\mu | \mathrm{vac} \rangle \neq 0$.\footnote{
    In case $\vec l_\mu^{(0)} \langle \mathrm{vac} | O^\mu | \mathrm{vac} \rangle = 0$, $\vec l_\mu^{(\rho)} \langle \mathrm{vac} | O^\mu | \mathrm{vac} \rangle = 0$ must be fulfilled element-wise.
		For an odd number of indices $\vec l_\mu^{(0)}$ does not exist, thus $\vec l_\mu^{(\rho_1)} = \vec l_\mu^{(1)}$ and the arguments for the orthogonal decomposition apply.}
Consequently, the vector of condensates $\vec l^{(\rho)}_\mu\langle \mathrm{vac} | O^\mu | \mathrm{vac} \rangle$ must satisfy
\begin{equation}\label{eq:vacconstr_nonort3}
    \vec l^{(\rho)}_\mu\langle \mathrm{vac} | O^\mu | \mathrm{vac} \rangle
    \in \mathrm{ker}
        \left[
        L_\rho^{-1} -
        \left(\begin{array}{cc} L_0^{-1} & 0 \\ 0 & 0 \end{array}\right) \right]
\end{equation}
for all operators $O_\mu$.
Note that, in general, the kernel of the matrix is degenerate.
Thus, the imposed vacuum constraints are not unique in the sense that \eqref{eq:vacconstr_nonort2} is the strongest constraint which can be deduced.
In particular because of the non-orthogonality of $\vec l^{(0)}_\mu$ and $\vec l^{(\rho_1)}_\mu$, the matrix in \eqref{eq:vacconstr_nonort2} has no block-diagonal form and possible interrelations among the additional medium terms remain hidden.

Because the matrix in \eqref{eq:vacconstr_nonort2} must have a non-trivial kernel, thus being not invertible, there is no matrix $\bar L$ with
\begin{equation}
    \bar L^{-1} = \left[
        L_\rho^{-1} -
        \left(\begin{array}{cc} L_0^{-1} & 0 \\ 0 & 0 \end{array}\right) \right] \: .
\end{equation}
This is merely another formulation of the problem that the full decomposition has to be reevaluated in case additional tensors enter the projection vector although the vacuum decomposition is known.

For the orthogonal decomposition, the vacuum limit reads
\begin{align}
	0 = \langle \mathrm{vac} | O_\mu | \mathrm{vac} \rangle_1
	= \vec l^{(1)}_\mu \cdot \vec a_{1}
    \: .
\end{align}
As the components of $\vec l^{(1)}_\mu$ must be linearly independent and the kernel of the invertible medium-specific projection matrix trivial, i.\,e.\ \(\ker \left( \vec l^{(1)} \circ \vec l^{(1)} \right)^{-1} = \{0\}\), this is equivalent to $\vec a_{1} \to 0$ with element-wise vanishing of the medium-specific condensates
\begin{align}
	\label{eq:vacconstrortho}
	0 = \vec l^{(1)}_\nu \langle \mathrm{vac} | O^\nu | \mathrm{vac} \rangle \: .
\end{align}
From \eqref{eq:medprojvec} it can be seen that medium-specific condensates $\vec l^{(1)}_{\mu} \langle\!\langle O^\mu \rangle\!\rangle$ contain algebraic vacuum condensates $\vec l^{(0)}_{\mu} \langle\!\langle O^\mu \rangle\!\rangle$.
Accordingly, the requirement of vanishing medium-specific condensates gives rise to vacuum constraints as interrelations among the \(\vec l^{(\rho)\mu} \langle \mathrm{vac} | O_\mu | \mathrm{vac} \rangle\) entering the medium-specific part of the decomposition, in particular also between terms which already occur in vacuum, i.\,e.\ \(\vec l^{(0)}_{\mu} \langle \mathrm{vac} | O^\mu | \mathrm{vac} \rangle\).
Thus, the vacuum limit of non-vacuum condensates \(\vec l^{(\rho_1)}_{\mu} \langle \mathrm{vac} | O^\mu | \mathrm{vac} \rangle\) is restricted by vacuum condensates:
\begin{align}\label{eq:vacconstr}
	\vec l^{(\rho_1)}_\mu \langle \mathrm{vac} | O^\mu | \mathrm{vac} \rangle = L_{\rho_1,0} L_0^{-1} \vec l^{(0)}_\mu \langle \mathrm{vac} | O^\mu | \mathrm{vac} \rangle \: .
\end{align}
In particular, non-vacuum condensates have a non-zero vacuum limit.

Summarizing, the Lorentz tensors which enter the decomposition upon the onset of a continuous parameter $\rho$, as e.\,g.\ the density, lead to additional condensates.
The limit of these condensates for $\rho \to 0$ is however constrained by \eqref{eq:vacconstr} if the transition is assumed to be continuous.
It is remarkable that these vacuum constraints are independent of the operator $O_\mu$, but only depend on the rank of this Lorentz tensor.
Finally, the arguments given above are not restricted to the occurrence of a medium velocity $v_\mu$ and can in principle be applied to any decomposition where additional tensors enter.

\subsection{Transformation to canonical condensates}

Utilizing quark and gluon equations of motion as well as Dirac matrix identities and symmetries of the QCD ground state, the elements of the vector of condensates can be mapped to a smaller set of (canonical) condensates,
\begin{align}
	\vec l^{(\rho)}_\mu \langle O^\mu \rangle = K \vec{\langle q \rangle} \: ,
\end{align}
where $K$ is a linear transformation matrix.
It maps the canonical condensates $\vec{\langle q \rangle} = (\vec{\langle q \rangle}{}^{(0)},\vec{\langle q \rangle}{}^{(\rho_1)})^\mathrm{T}$ onto the contracted \glspl{EV} $\vec l^{(\rho)}_\mu \langle O^\mu \rangle$.
Canonical condensates contain the minimal number of covariant derivatives.
Whereas the Lorentz decomposition of a QCD operator depends only on its tensor rank, the operator itself might be identified by its transformation to canonical condensates.
Having fixed a set of canonical condensates in the medium, any operator $O$ may be specified by its matrix $K$.

Since vacuum and non-vacuum parts of the ($m \times n$) matrix $K$ separate according to
\begin{align}
	K = \left(\begin{array}{cc}
						K_0 & 0\\
						  0 & K_{\rho_1}
						\end{array}\right) \: 
\end{align}
the vacuum constraints \eqref{eq:vacconstr} read
\begin{align}
	K_{\rho_1} \vec{\langle q \rangle}{}^{(\rho_1)} = L_{\rho_1,0} L_0^{-1} K_0 \vec{\langle q \rangle}{}^{(0)} \: ,
\end{align}
requiring the left inverse of $K_{\rho_1}$ to cast the vacuum constraints in the desired form. Since columns and rows in $K$ might show linear dependencies the inverse $K_{\rho_1}^{-1}$ does not exist in general. However, a left inverse $K_{\rho_1}^+$ can always be constructed from a matrix with linearly independent columns which can be obtained by appropriate redefinition of $\vec{\langle q \rangle}$.
Then the vacuum constraints of canonical condensates read
\begin{align}
	\vec{\langle q \rangle}{}^{(\rho_1)} = K_{\rho_1}^+ L_{\rho_1,0} L_0^{-1} K_0 \vec{\langle q \rangle}{}^{(0)} \: .
\end{align}

\section{Example: four-quark condensates}
\label{sec:4q}

The above derived formulae are essential for setting up in-medium \glspl{QSR} for mesons and baryons.
Depending on the order of the \gls{OPE} of the correlator, operators with an increasing number of Lorenz indices occur.
To illustrate the definitions of vacuum and medium-specific contributions and to shed light on the effect of vacuum constraints we choose four-quark condensates as an example with up to five Lorentz indices.
The evaluation of leading order $\alpha_\mathrm{s}$ four-quark condensate terms using the Fock-Schwinger gauge method \cite{Novikov:1983gd,Pascual:1984zb} requires the computation of three distinct contributions to the meson current-current correlator (cf.\ Appendix of Ref.~\cite{Buchheim:2014rpa}):
$\langle\!\langle \bar{q} \overset{{}_\leftarrow}{D}_\mu \overset{{}_\leftarrow}{D}_\nu \overset{{}_\leftarrow}{D}_\lambda \Gamma \rangle\!\rangle$, $\langle\!\langle \bar{q} \Gamma [D_\nu, G_{\kappa\lambda}] q \rangle\!\rangle$, $\langle\!\langle \bar{q} \overset{{}_\leftarrow}{D}_\mu \Gamma G_{\nu\lambda} q \rangle\!\rangle$.
Specifying the Dirac structure $\Gamma$ for a light vector current, such as for the $\uprho$ or $\omega$ meson, we end up with the following Gibbs averaged operators:\\

\begin{minipage}[t]{0.47\textwidth}
	even \gls{OPE}
	\begin{enumerate}[label={\itshape (e\arabic{*})}, ref={\itshape (e\arabic{*})}, itemsep=0.0ex]
		\item $\langle\!\langle \bar{q} \overset{{}_\leftarrow}{D}_\mu \overset{{}_\leftarrow}{D}_\nu \overset{{}_\leftarrow}{D}_\lambda \gamma_\alpha q \rangle\!\rangle$ \label{enum:e1}
		\item $\langle\!\langle \bar{q} \gamma_\mu [D_\nu, G_{\kappa\lambda}] q \rangle\!\rangle$ \label{enum:e2}
		\item $\langle\!\langle \bar{q} \gamma_5 \gamma_\mu [D_\nu, G_{\kappa\lambda}] q \rangle\!\rangle\!\rangle$ \label{enum:e3}
		\item $\langle\!\langle \bar{q} \overset{{}_\leftarrow}{D}_\mu \gamma_\alpha G_{\nu\lambda} q \rangle\!\rangle$ \label{enum:e4}
		\item $\langle\!\langle \bar{q} \overset{{}_\leftarrow}{D}_\mu \gamma_5 \gamma_\alpha G_{\nu\lambda} q \rangle\!\rangle$ \label{enum:e5}
	\end{enumerate}
\end{minipage}
\hfill
\begin{minipage}[t]{0.47\textwidth}
	odd \gls{OPE}
	\begin{enumerate}[label={\itshape (o\arabic{*})}, ref={\itshape (o\arabic{*})}, itemsep=0.0ex]
		\item $\langle\!\langle \bar{q} \overset{{}_\leftarrow}{D}_\mu \overset{{}_\leftarrow}{D}_\nu \overset{{}_\leftarrow}{D}_\lambda q \rangle\!\rangle$ \label{enum:o1}
		\item $\langle\!\langle \bar{q} [D_\nu, G_{\kappa\lambda}] q \rangle\!\rangle$ \label{enum:o2}
		\item $\langle\!\langle \bar{q} \overset{{}_\leftarrow}{D}_\mu G_{\nu\lambda} q \rangle\!\rangle$ \label{enum:o3}
		\item $\langle\!\langle \bar{q} \overset{{}_\leftarrow}{D}_\mu G_{\nu\lambda} q \rangle\!\rangle$ \label{enum:o4}
		\item $\langle\!\langle \bar{q} \overset{{}_\leftarrow}{D}_\mu \sigma_{\alpha\beta} G_{\nu\lambda} q \rangle\!\rangle$ \label{enum:o5}
	\end{enumerate}
\end{minipage}\\[1.5ex]

\noindent and equivalent objects with covariant derivatives acting on the right quark operator, i.\,e.\ $\bar q \overset{{}_\leftarrow}{D}_\mu \cdots q \longrightarrow \bar q \cdots D_\mu q$.
Four-quark condensates in leading order $\alpha_\mathrm{s}$ also occur in the next-to-leading order correlator with inserted interaction term:\\

\begin{minipage}[t]{0.47\textwidth}
	even \gls{OPE}
	\begin{enumerate}[label={\itshape (e\arabic{*})}, ref={\itshape (e\arabic{*})}, itemsep=0.0ex, start=6]
		\item $\langle\!\langle \bar{q} \gamma_\mu T^A q \bar{q}' \gamma_\nu T^A q' \rangle\!\rangle$ \label{enum:e6}
		\item $\langle\!\langle \bar{q} \gamma_5 \gamma_\mu T^A q \bar{q}' \gamma_5 \gamma_\nu T^A q' \rangle\!\rangle$ \label{enum:e7}
	\end{enumerate}
\end{minipage}
\hfill
\begin{minipage}[t]{0.47\textwidth}
	odd \gls{OPE}
	\begin{enumerate}[label={\itshape (o\arabic{*})}, ref={\itshape (o\arabic{*})}, itemsep=0.0ex, start=6]
		\item $\langle\!\langle \bar{q} T^A q \bar{q}' \gamma_\mu T^A q' \rangle\!\rangle$ \label{enum:o6}
		\item $\langle\!\langle \bar{q} \gamma_5 \gamma_\mu T^A q \bar{q}' \sigma_{\nu\lambda} T^A q' \rangle\!\rangle$ \label{enum:o7}
	\end{enumerate}
\end{minipage}\\[1.5ex]
where $q$ and $q'$ denote light-quark flavours, which may coincide, and $T^A$ with $A=0,\ldots,8$ symbolize the generators of the colour group $\mathrm{SU}(3)_\mathrm{c}$, $t^A$, added by the unit element for $A=0$.

In the following we provide vacuum and medium-specific decompositions of those \gls{OPE} operators \ref{enum:e1}--\ref{enum:e7} and \ref{enum:o1}--\ref{enum:o7} which do not vanish in the light (axial-)vector meson \gls{OPE} (due to (anti)symmetries among the Lorenz-indices or vanishing Dirac-trace results).
The Lorentz-contracted operators occurring in the decomposition coefficients are transformed to canonical condensates.
The corresponding transformation matrices are presented along with the vacuum constraints in terms of these canonical condensates.

In pseudo-scalar, scalar, vector and axial-vector meson currents the Dirac trace results of the \gls{OPE} differ only by a constant (non-zero) factor/sign, except for traces with Dirac projection matrix $\sigma_{\alpha\beta}$ where vector and axial-vector currents give zero results.
Hence, up to prefactors all four-quark condensate contributions in leading order $\alpha_\mathrm{s}$ to spin-1 \glspl{OPE} already enter spin-0 \glspl{OPE}.
The presented constraints are therefore valid in both channels.
In particular, they do not contradict.

\subsection{Even OPE}
\label{subsec:evenOPE}

In order to extend conveniently the following results to mesons consisting of a light and a heavy quark, such as $D$ and $B$ mesons, we first present the vacuum constraints set up by the contributions of \ref{enum:e1}--\ref{enum:e5} and include additional vacuum constraints from \ref{enum:e6} and \ref{enum:e7} originating from the next-to-leading order correlator later on.

\begin{enumerate}[label={\itshape (\roman{*})}, ref={\itshape (\roman{*})}, itemindent=2.5em, leftmargin=0.0em]
	\label{enum:even1}
	\item The first \gls{EV} \ref{enum:e1}
 decomposes into vacuum and medium-specific terms as (cf.\ \eqref{eq:defmed})
\begin{align}
& \langle\!\langle \bar{q} \overset{{}_\leftarrow}{D}_\mu \overset{{}_\leftarrow}{D}_\nu \overset{{}_\leftarrow}{D}_\lambda \gamma_\alpha q \rangle\!\rangle_0 \nonumber\\
& = 
	\frac{1}{72}
	\left(\begin{array}{l}
					g_{\mu\lambda} g_{\nu\alpha} + g_{\mu\alpha} g_{\nu\lambda}\\
					g_{\mu\nu} g_{\lambda\alpha}
				\end{array}\right)^\mathrm{T}
	\left(\begin{array}{cc}
					2 & -1\\
					-1 & 5
				\end{array}\right)
	\left(\begin{array}{l}
					\langle\!\langle \bar{q} \overset{{}_\leftarrow}{D}{}^\mu \overset{{}_\leftarrow}{\slashed{D}} \overset{{}_\leftarrow}{D}_\mu q + \bar{q} \overset{{}_\leftarrow}{\slashed{D}} \overset{{}_\leftarrow}{D}{}^2 q \rangle\!\rangle\\
					\langle\!\langle \bar{q} \overset{{}_\leftarrow}{D}{}^2 \overset{{}_\leftarrow}{\slashed{D}} q \rangle\!\rangle
				\end{array}\right) \: , \\
& \langle\!\langle \bar{q} \overset{{}_\leftarrow}{D}_\mu \overset{{}_\leftarrow}{D}_\nu \overset{{}_\leftarrow}{D}_\lambda \gamma_\alpha q \rangle\!\rangle_1 \nonumber\\
& = \frac{1}{240}
	\left(\begin{array}{l}
					g_{\mu\lambda} g_{\nu\alpha} + g_{\mu\alpha} g_{\nu\lambda} - \frac{2}{v^2}( g_{\mu\lambda} v_\nu v_\alpha\\
					\qquad + g_{\mu\alpha} v_\nu v_\lambda + g_{\nu\alpha} v_\mu v_\lambda + g_{\nu\lambda} v_\mu v_\alpha ) \\
					g_{\mu\nu} g_{\lambda\alpha} - \frac{4}{v^2} g_{\mu\nu} v_\lambda v_\alpha \\
					g_{\mu\nu} g_{\lambda\alpha} - \frac{4}{v^2} g_{\lambda\alpha} v_\mu v_\nu \\
					g_{\mu\lambda} g_{\nu\alpha} + g_{\mu\alpha} g_{\nu\lambda} + g_{\mu\nu} g_{\lambda\alpha} -\frac{48}{v^4} v_\mu v_\nu v_\lambda v_\alpha
				\end{array}\right)^\mathrm{T}
	\left(\begin{array}{cccc}
					8 & -1 & -1 & -2\\
					-1 & 7 & 1 & -1\\
					-1 & 1 & 7 & -1\\
					-2 & -1& -1 & \frac{4}{3}
				\end{array}\right)
	\nonumber\\
	&\quad\times
	\left(\begin{array}{l}
					\langle\!\langle \bar{q} \overset{{}_\leftarrow}{D}{}^{\mu'} \overset{{}_\leftarrow}{\slashed{D}} \overset{{}_\leftarrow}{D}_{\mu'} q + \bar{q} \overset{{}_\leftarrow}{\slashed{D}} \overset{{}_\leftarrow}{D}{}^2 q - \frac{2}{v^2} ( \bar{q} \overset{{}_\leftarrow}{D}{}^{\mu'} (v\overset{{}_\leftarrow}{D}) \overset{{}_\leftarrow}{D}_{\mu'} \slashed{v} q \\
					\qquad + \bar{q} \overset{{}_\leftarrow}{\slashed{D}} (v\overset{{}_\leftarrow}{D})^2 q + \bar{q} (v\overset{{}_\leftarrow}{D}) \overset{{}_\leftarrow}{\slashed{D}} (\overset{{}_\leftarrow}{D}) q + \bar{q} (v\overset{{}_\leftarrow}{D}) \overset{{}_\leftarrow}{D}{}^2 \slashed{v} q ) \rangle\!\rangle \\
					\langle\!\langle \bar{q} \overset{{}_\leftarrow}{D}{}^2 \overset{{}_\leftarrow}{\slashed{D}} q - \frac{4}{v^2} \bar{q} \overset{{}_\leftarrow}{D}{}^2 (\overset{{}_\leftarrow}{D}) \slashed{v} q \rangle\!\rangle \\
					\langle\!\langle \bar{q} \overset{{}_\leftarrow}{D}{}^2 \overset{{}_\leftarrow}{\slashed{D}} q - \frac{4}{v^2} \bar{q} (v\overset{{}_\leftarrow}{D})^2 \overset{{}_\leftarrow}{\slashed{D}} q \rangle\!\rangle\\
					\langle\!\langle \bar{q} \overset{{}_\leftarrow}{D}{}^{\mu'} \overset{{}_\leftarrow}{\slashed{D}} \overset{{}_\leftarrow}{D}_{\mu'} q + \bar{q} \overset{{}_\leftarrow}{\slashed{D}} \overset{{}_\leftarrow}{D}{}^2 q + \bar{q} \overset{{}_\leftarrow}{D}{}^2 \overset{{}_\leftarrow}{\slashed{D}} q - \frac{48}{v^4} \bar{q} (v\overset{{}_\leftarrow}{D})^3 \slashed{v} q \rangle\!\rangle
				\end{array}\right) \: ,
	\label{eq:MedSpecCond1}
\end{align}
where only the part symmetric in the index pairs $\mu\nu$ and $\lambda\alpha$ contributes to the (axial-)vector \gls{OPE}.
Transformation to canonical condensates yields
\begin{align}
	\left(\!\!\!\begin{array}{l}
		\langle \bar{q} \overset{{}_\leftarrow}{D}{}^2 \overset{{}_\leftarrow}{\slashed{D}} q \rangle\\
		\langle \bar{q} \overset{{}_\leftarrow}{D}{}^{\mu'} \overset{{}_\leftarrow}{\slashed{D}} \overset{{}_\leftarrow}{D}_{\mu'} q \rangle\\
		\langle \bar{q} \overset{{}_\leftarrow}{\slashed{D}} \overset{{}_\leftarrow}{D}{}^2 q \rangle\\[1.5ex]
		\langle \bar{q} \overset{{}_\leftarrow}{D}{}^2 (v\overset{{}_\leftarrow}{D}) \slashed{v} q \rangle\\
		\langle \bar{q} \overset{{}_\leftarrow}{D}{}^{\mu} (v\overset{{}_\leftarrow}{D}) \overset{{}_\leftarrow}{D}_{\mu} \slashed{v} q \rangle\\
		\langle \bar{q} \overset{{}_\leftarrow}{\slashed{D}} (v\overset{{}_\leftarrow}{D})^2 q \rangle\\
		\langle \bar{q} (v\overset{{}_\leftarrow}{D})^2 \overset{{}_\leftarrow}{\slashed{D}} q \rangle\\
		\langle \bar{q} (v\overset{{}_\leftarrow}{D}) \overset{{}_\leftarrow}{\slashed{D}} (v\overset{{}_\leftarrow}{D}) q \rangle\\
		\langle \bar{q} (v\overset{{}_\leftarrow}{D}) \overset{{}_\leftarrow}{D}{}^2 \slashed{v} q \rangle\\
		\langle \bar{q} (v\overset{{}_\leftarrow}{D})^3 \slashed{v} q \rangle
	\end{array}\!\!\!\right)
	=
	\left(\begin{array}{ccccccc}
		0           & i &&&&&\\
		\frac{i}{2} & i &&&&&\\
		0           & i &&&&&\\
		&& 0 					 & 1 & 0 & 0   & 0\\
		&& \frac{i}{2} & 1 & 0 & 0   & 0\\
		&& 0           & 0 & i & 0   & 0\\
		&& 0           & 0 & i & 0   & 0\\
		&& 0           & 0 & i & -i  & 0\\
		&& 0           & 1 & 0 & -2i & 0\\
		&& 0           & 0 & 0 & 0   & 1
	\end{array}\right)
	\left(\!\!\begin{array}{l}
		g^2 \langle \bar q \gamma^\mu t^A q \sum_f \bar f \gamma_\mu t^A f \rangle\\
		\frac{m_q}{2} \langle \bar q g \sigma G q \rangle  - m_q^3 \langle \bar q q \rangle\\[1.5ex]
		g^2 \langle \bar q \slashed{v} t^A q \sum_f \bar f \slashed{v} t^A f \rangle\\
		\frac{1}{2}\langle \bar q (v\overset{{}_\leftarrow}{D}) g \sigma G \slashed{v} q \rangle\\
				\qquad\quad - m_q^2 \langle \bar q (v\overset{{}_\leftarrow}{D}) \slashed{v} q \rangle\\
		m_q \langle \bar q (v\overset{{}_\leftarrow}{D})^2 q \rangle\\
		g \langle \bar q (v\overset{{}_\leftarrow}{D}) \gamma^\mu G_{\mu\nu} v^\nu q \rangle\\
		\langle \bar q (v\overset{{}_\leftarrow}{D})^3 \slashed{v} q \rangle
	\end{array}\!\!\!\right) \: ,
\end{align}
where $\sum_f$ denotes the sum over light-quark flavours.
The decomposition of the terms \ref{enum:e2}--\ref{enum:e5} proceeds analogously. The results along with the associated canonical condensates are listed in Appendix~\ref{app:furtherdecompeven}.
\end{enumerate}

Gathering the vacuum constraints as vanishing medium-specific condensates of the decompositions \eqref{eq:MedSpecCond1}, \eqref{eq:MedSpecCond2}, \eqref{eq:MedSpecCond3} and \eqref{eq:MedSpecCond4} in terms of canonical condensates yields
\begin{subequations}
\label{eq:VacContCanon}
\begin{align}
	0 & = \langle \mathrm{vac} | g^2 \bar{q} \gamma^\mu t^A q \sum_f \bar f \gamma_\mu t^A f + 2 m_q \bar{q} g \sigma G q - 4 m_q^3 \bar{q} q - \frac{2}{v^2} \bigg( g^2 \bar{q} \slashed{v} t^A q \sum_f \bar f \slashed{v} t^A f 
	\label{eq:VacContCanonFirst}\nonumber\\
	& - 2 \bar{q} (vi\overset{{}_\leftarrow}{D}) g \sigma G \slashed{v} q + 4 m_q^2 \bar{q} (vi\overset{{}_\leftarrow}{D}) \slashed{v} q - 4 m_q \bar{q} (vi\overset{{}_\leftarrow}{D})^2 q + 6 g \bar{q} (v\overset{{}_\leftarrow}{D}) \gamma^\mu G_{\nu\mu} v^\nu q \bigg) | \mathrm{vac} \rangle \: , \\
	0 & = \langle \mathrm{vac} | m_q \bar{q} g \sigma G q - 2 m_q^3 \bar{q} q -\frac{4}{v^2} \left( - \bar{q} (vi\overset{{}_\leftarrow}{D}) g \sigma G \slashed{v} g + 2m_q^2 \bar{q} (vi\overset{{}_\leftarrow}{D}) \slashed{v} q \right) | \mathrm{vac} \rangle \: , \\
	0 & = \langle \mathrm{vac} | m_q \bar{q} g \sigma G q - 2 m_q^3 \bar{q} q + \frac{8}{v^2} m_q \bar{q} (vi\overset{{}_\leftarrow}{D})^2 q | \mathrm{vac} \rangle \: , \\
	0 & = \langle \mathrm{vac} | g^2 \bar{q} \gamma^\mu t^A q \sum_f \bar f \gamma_\mu t^A f + 3 m_q \bar{q} g \sigma G q - 6 m_q^3 \bar{q} q - \frac{48}{v^4} \bar{q} (vi\overset{{}_\leftarrow}{D})^3 \slashed{v} q | \mathrm{vac} \rangle \: , \\[1ex]
	0 & = \langle \mathrm{vac} |  g \bar{q} \gamma^\mu t^A q \sum_f \bar f \gamma_\mu t^A f - \frac{2}{v^2} \bigg( \bar{q} \gamma^\mu [(vD),G_{\mu\nu}] v^\nu q +  g \bar{q} \slashed{v} t^A q \sum_f \bar f \slashed{v} t^A f \bigg) | \mathrm{vac} \rangle \: , \\
	0 & = \langle \mathrm{vac} | g \bar{q} \slashed{v} t^A q \sum_f \bar f \slashed{v} t^A f + \bar{q} \gamma^\mu [(vD),G_{\mu\nu}] v^\nu q + \bar{q} \slashed{v} \sigma^{\nu\lambda} [(vD),G_{\nu\lambda}] q | \mathrm{vac} \rangle \: , \\
	0 & = \langle \mathrm{vac} | 3 g^2 \bar{q} \gamma^\mu t^A q \sum_f \bar f \gamma_\mu t^A f + 3 m_q \bar{q} g \sigma G q - \frac{4}{v^2} \bigg( -q \varepsilon_{\mu\nu\lambda\tau} v^\tau \bar{q} i\overset{{}_\leftarrow}{D}{}^\mu \gamma_5 \slashed{v} G^{\nu\lambda} q \nonumber\\
	& - 2g m_q \bar{q} v_\mu \sigma{\mu\nu} G_{\nu\lambda} v^\lambda q + g^2 \bar{q} \slashed{v} t^A q \sum_f \bar f \slashed{v} t^A f + 2g \bar{q} (v\overset{{}_\leftarrow}{D}) \gamma^\lambda G_{\nu\lambda} v^\nu q \bigg)| \mathrm{vac} \rangle
	\label{eq:VacContCanonLast}
    \: .
\end{align}
\end{subequations}

Equations \eqref{eq:VacContCanon} relate the known \glspl{VEV} of vacuum condensates, e.\,g., $\langle \mathrm{vac} | \bar q q | \mathrm{vac} \rangle$, $g \langle \mathrm{vac} | \bar q \sigma G q | \mathrm{vac} \rangle$ and $g^2 \langle \mathrm{vac} | \bar q \gamma^\mu t^A q \sum_f \bar f \gamma_\mu t^A f | \mathrm{vac} \rangle$, or the known components of medium-specific condensates up to mass dimension 5, i.\,e.\ $\langle \mathrm{vac} | \bar q (vi\overset{{}_\leftarrow}{D}) \slashed{v} q | \mathrm{vac} \rangle$ and $\langle \mathrm{vac} | \bar q (vi\overset{{}_\leftarrow}{D})^2 q | \mathrm{vac} \rangle$, to heretofore unknown \glspl{VEV} of components of medium-specific condensates in mass dimension 6, e.\,g.\ the medium four-quark condensate $g^2 \langle \mathrm{vac} | \bar q \slashed{v} t^A q \sum_f \bar f \slashed{v} t^A f | \mathrm{vac} \rangle/v^2$.
We therefore solve the above system of linear equations for the unknown condensates.

The system of equations is underdetermined, thus, the solution is not unique.
In fact, the relation between vacuum and medium four-quark condensate can be tuned this way.
The \glspl{VEV} of non-vacuum operators of mass dimension 6 deduced from the algebraic vacuum constraints, where we choose
\begin{subequations}
\begin{align}
	& g^2 \langle \mathrm{vac} | \bar{q} \slashed{v} t^A q \sum_f \bar f \slashed{v} t^A f | \mathrm{vac} \rangle / v^2 =
	x_q
	\label{eq:relEven1} \: , \\
	& g \langle \mathrm{vac} | m_q \bar{q} v_\mu \sigma^{\mu\nu} G_{\nu\lambda} v^\lambda q | \mathrm{vac} \rangle / v^2 = y_q \: ,
\end{align}
\end{subequations}
to be the free parameters of the solution of the system of equations \eqref{eq:VacContCanon}, read
\begin{subequations}
\label{eq:evenAVLsoft}
\begin{align}
	& g \langle \mathrm{vac} | \bar{q} (vi\overset{{}_\leftarrow}{D}) \sigma G \slashed{v} q | \mathrm{vac} \rangle / v^2 =
	\frac{1}{4} \bigg( 4 m_q^3 \langle \mathrm{vac} | \bar q q | \mathrm{vac} \rangle - g m_q \langle \mathrm{vac} | \bar q \sigma G q | \mathrm{vac} \rangle \bigg) \: , \\
	& g \langle \mathrm{vac} | \bar{q} (vi\overset{{}_\leftarrow}{D}) \gamma^\mu G_{\nu\mu} v^\nu q | \mathrm{vac} \rangle / v^2 =
	\frac{1}{12} \bigg( g^2 \langle \mathrm{vac} | \bar q \gamma^\mu t^A q \sum_f \bar f \gamma_\mu t^A f | \mathrm{vac} \rangle - 2 x_q \bigg) \: , 
\end{align}
\begin{align}
	& g \langle \mathrm{vac} | \bar{q} \gamma^\mu [(vD),G_{\mu\nu}] v^\nu q | \mathrm{vac} \rangle / v^2 =
	\frac{1}{2} \bigg( g^2 \langle \mathrm{vac} | \bar{q} \gamma^\mu t^A q \sum_f \bar f \gamma_\mu t^A f | \mathrm{vac} \rangle - 2 x_q \bigg) \: , \\
	& g \langle \mathrm{vac} | \bar{q} \slashed{v} \sigma^{\nu\lambda} [(viD),G_{\nu\lambda}] q | \mathrm{vac} \rangle / v^2 =
	\frac{1}{2} \bigg( g^2 \langle \mathrm{vac} | \bar{q} \gamma^\mu t^A q \sum_f \bar f \gamma_\mu t^A f | \mathrm{vac} \rangle - 4 x_q \bigg) \: , 
	\label{eq:zeroavl}\\
	& g \langle \mathrm{vac} | \varepsilon_{\mu\nu\lambda\tau} v^\tau \bar{q} i \overset{{}_\leftarrow}{D}{}^\mu \gamma_5 \slashed{v} G^{\nu\lambda} q | \mathrm{vac} \rangle / v^2 = 
	\frac{1}{12} \bigg( 6 m_q^3 \langle \mathrm{vac} | \bar q q | \mathrm{vac} \rangle - 8 g m_q \langle \mathrm{vac} | \bar q \sigma G q | \mathrm{vac} \rangle \nonumber\\
	& \quad - 7 g^2 \langle \mathrm{vac} | \bar{q} \gamma^\mu t^A q \sum_f \bar f \gamma_\mu t^A f | \mathrm{vac} \rangle + 8 x_q - 24 y_q \bigg) \: , \\
	& \langle \mathrm{vac} | \bar{q} (vi\overset{{}_\leftarrow}{D})^3 \slashed{v} q | \mathrm{vac} \rangle / v^4 =
	\frac{1}{48} \bigg( -6 m_q^3 \langle \mathrm{vac} | \bar q q | \mathrm{vac} \rangle + 3 g m_q \langle \mathrm{vac} | \bar q \sigma G q | \mathrm{vac} \rangle \nonumber\\
	& \quad + g^2 \langle \mathrm{vac} | \bar{q} \gamma^\mu t^A q \sum_f \bar f \gamma_\mu t^A f | \mathrm{vac} \rangle \bigg) \: ,
\end{align}
\end{subequations}
where the vacuum constraints
\begin{align}
	& \langle \mathrm{vac} | \bar q(vi\overset{{}_\leftarrow}{D}) \slashed{v} q | \mathrm{vac} \rangle / v^2 = -\frac{1}{4} m_q \langle \mathrm{vac} | \bar q q | \mathrm{vac} \rangle
	\label{eq:evenDim4AVL} \: , \\
	& \langle \mathrm{vac} | \bar{q} (vi\overset{{}_\leftarrow}{D})^2 q | \mathrm{vac} \rangle / v^2 = - \frac{1}{8} g \langle \mathrm{vac} | \bar q \sigma G q | \mathrm{vac} \rangle + \frac{1}{4} m_q^2 \langle \mathrm{vac} | \bar q q | \mathrm{vac} \rangle
	\label{eq:evenDim5AVL}
\end{align}
from mass dimension 4 and 5 condensates have been used, respectively.
The same solution holds in the light chiral limit $m_q \rightarrow 0$.

\begin{enumerate}[label={\itshape (\roman{*})}, ref={\itshape (\roman{*})}, itemindent=2.5em, leftmargin=0.0em, start=2]
	\label{enum:even6}
	\item Gibbs averaged light four-quark operators \ref{enum:e6} and \ref{enum:e7} from the next-to-leading order correlator decompose into vacuum and medium-specific terms as
\begin{align}
	\langle\!\langle \bar q \Gamma' \gamma_\mu T^A q \bar q' \Gamma' \gamma_\nu T^A q' \rangle\!\rangle_0 & = \frac{1}{4}g_{\mu\nu} \langle\!\langle \bar q \Gamma' \gamma^{\mu'} T^A q \bar q' \Gamma' \gamma_{\mu'} T^A q' \rangle\!\rangle
	\label{eq:VacConNLOCorr1}\\
	\langle\!\langle \bar q \Gamma' \gamma_\mu T^A q \bar q' \Gamma' \gamma_\nu T^A q' \rangle\!\rangle_1 & = \frac{1}{12}\bigg( g_{\mu\nu} - \frac{4}{v^2}v_\mu v_\nu \bigg) \nonumber\\
	& \qquad\quad \times \langle\!\langle \bar q \Gamma' \gamma^{\mu'} T^A q \bar q' \Gamma' \gamma_{\mu'} T^A q' - \frac{4}{v^2} \bar q \Gamma' \slashed{v} T^A q \bar q' \Gamma' \slashed{v} T^A q' \rangle\!\rangle \: ,
	\label{eq:VacConNLOCorr2}
\end{align}
where  $\Gamma'$ denotes either $1$ or $\gamma_5$.
These condensates originating from two cut quark lines in the corresponding diagram of the next-to-leading order correlator are already in canonical form.
\end{enumerate}

The vacuum constraints from vanishing medium-specific condensates \eqref{eq:VacConNLOCorr2} read
\begin{align}
	\langle \mathrm{vac} | \bar q \Gamma' \slashed{v} T^A q \bar q' \Gamma' \slashed{v} T^A q' | \mathrm{vac} \rangle/v^2\ = \frac{1}{4} \langle \mathrm{vac} | \bar q \Gamma' \gamma^{\mu'} T^A q \bar q' \Gamma' \gamma_{\mu'} T^A q' | \mathrm{vac} \rangle
	\label{eq:SolVacConNLOCorr}
\end{align}
which completes the list of Eqs.~\eqref{eq:evenAVLsoft} for light quark condensates.
They confine the arbitrary parameter $x_q$ in \eqref{eq:evenAVLsoft}, i.\,e.\ Eq.~\eqref{eq:SolVacConNLOCorr} with $\Gamma'=1$ and $T^A=t^A$ determines the relation between light four-quark condensates in vacuum and medium, such that $x_q = \frac{1}{4} \langle \mathrm{vac} | \bar q \gamma^{\mu'} t^A q \sum_f \bar f \gamma_{\mu'} t^A f | \mathrm{vac} \rangle$ must be used in the above vacuum constraints \eqref{eq:evenAVLsoft} leading to a zero \gls{VEV} in \eqref{eq:zeroavl}.

This is a general result providing vacuum values for non-vacuum condensates.
It is deduced only from the requirement of a continuous transition to vacuum terms when "turning off the medium", $\rho \rightarrow 0$.

Vacuum constraints of the even \gls{OPE} of heavy-light meson currents can be deduced from the results presented in this subsection.
Besides Gibbs averaged light-quark operators also heavy-quark operators and such quantities containing both kinds of quark operators enter the \gls{OPE} of heavy-light mesons.
The Gibbs averages \ref{enum:e1}--\ref{enum:e5} are supplemented by similar contributions where $\bar q \overset{{}_\leftarrow}{D}_\mu \cdots q \longrightarrow \bar Q \cdots D_\mu Q$.
The four-quark terms \ref{enum:e6} and \ref{enum:e7} are substituted by similar contributions where one $\bar q \cdots q$ pair is changed for one $\bar Q \cdots Q$ pair.

The decompositions of the heavy-quark terms and their vacuum constraints originating from the leading order correlator can be obtained by $q \longrightarrow Q$ in \eqref{eq:MedSpecCond1}--\eqref{eq:evenDim5AVL}, where this substitution also refers to subscripts, e.\,g.\ further parameters enter the heavy-quark vacuum constraints, $x_Q$ and $y_Q$, which are absent in a light-quark current \gls{OPE}.
The decomposition of the heavy-light four-quark terms originating from the next-to-leading order correlator can be obtained by the exchange of one quark pair $\bar q \cdots q \longrightarrow \bar Q \cdots Q$ in \eqref{eq:VacConNLOCorr1}--\eqref{eq:SolVacConNLOCorr}.
This substitution in Eq.~\eqref{eq:SolVacConNLOCorr} fixes the parameter $x_Q = \frac{1}{4} \langle \mathrm{vac} | \bar Q \gamma^{\mu'} t^A Q \sum_f \bar f \gamma_{\mu'} t^A f | \mathrm{vac} \rangle$.

A subtlety may be mentioned regarding the parameter $x_q$ in heavy-light current \glspl{OPE}.
Since no light four-quark terms \ref{enum:e6} and \ref{enum:e7} but exclusively corresponding four-quark terms containing heavy and light quark pairs enter such \glspl{OPE}, $x_q$ can not be determined by the heavy-light current \gls{OPE} alone.
However, universality of condensates requires identical behavior of identical condensates in different \glspl{OPE}.
Thus, the light current \gls{OPE} fixes the parameter $x_q$ also in a heavy-light current \gls{OPE}.

\subsection{Odd OPE}
\label{subsec:oddOPE}

In vacuum, the Lorentz-odd \gls{OPE} is always zero.
Hence, $\langle\!\langle O \rangle\!\rangle_0 = 0$ for all Lorentz-odd operators.
\begin{enumerate}[label={\itshape (\roman{*})}, ref={\itshape (\roman{*})}, itemindent=2.5em, leftmargin=0.0em]
	\label{enum:odd1}
	\item The first \gls{EV} \ref{enum:o1} decomposes as
\begin{align}
	&\langle\!\langle \bar{q} \overset{{}_\leftarrow}{D}_\mu \overset{{}_\leftarrow}{D}_\nu \overset{{}_\leftarrow}{D}_\lambda q \rangle\!\rangle = 
	\frac{1}{3v^2}
	\left(\!\!\begin{array}{l}
					v_\mu g_{\nu\lambda} \\
					v_\nu g_{\mu\lambda} \\
					v_\lambda g_{\mu\nu} \\
					\frac{1}{v^2} v_\mu v_\nu v_\lambda
				\end{array}\!\!\right)^\mathrm{T}
	\left(\begin{array}{cccc}
					 1 &  0 &  0 & -1\\
					 0 &  1 &  0 & -1\\
					 0 &  0 &  1 & -1\\
					-1 & -1 & -1 &  6
				\end{array}\right)
	\left(\!\!\!\begin{array}{l}
					\langle\!\langle \bar{q} (v\overset{{}_\leftarrow}{D}) \overset{{}_\leftarrow}{D}{}^2q \rangle\!\rangle\\
					\langle\!\langle \bar{q} \overset{{}_\leftarrow}{D}{}^{\mu'} (v\overset{{}_\leftarrow}{D}) \overset{{}_\leftarrow}{D}_{\mu'} q \rangle\!\rangle\\
					\langle\!\langle \bar{q} \overset{{}_\leftarrow}{D}{}^2 (v\overset{{}_\leftarrow}{D}) q \rangle\!\rangle\\
					\frac{1}{v^2} \langle\!\langle \bar{q} (v\overset{{}_\leftarrow}{D})^3 q \rangle\!\rangle 
				\end{array}\!\!\!\right) \: ,
	\label{eq:MedSpecCondOdd1}
\end{align}
where only the combination $\langle\!\langle \bar{q} (v\overset{{}_\leftarrow}{D}) \overset{{}_\leftarrow}{D}{}^2q + 2 \bar{q} \overset{{}_\leftarrow}{D}{}^{\mu'} (v\overset{{}_\leftarrow}{D}) \overset{{}_\leftarrow}{D}_{\mu'} q + \bar{q} \overset{{}_\leftarrow}{D}{}^2 (v\overset{{}_\leftarrow}{D}) q \rangle\!\rangle$ as well as the single term $\langle\!\langle \bar{q} (v\overset{{}_\leftarrow}{D})^3 q \rangle\!\rangle / v^2$ enter the (axial-)vector \gls{OPE}.
Transformation to canonical condensates yields
\begin{align}
	\left(\begin{array}{l}
					\langle \bar{q} (v\overset{{}_\leftarrow}{D}) \overset{{}_\leftarrow}{D}{}^2q \rangle\\
					\langle \bar{q} \overset{{}_\leftarrow}{D}{}^{\mu'} (v\overset{{}_\leftarrow}{D}) \overset{{}_\leftarrow}{D}_{\mu'} q \rangle\\
					\langle \bar{q} \overset{{}_\leftarrow}{D}{}^2 (v\overset{{}_\leftarrow}{D}) q \rangle\\
					\frac{1}{v^2} \langle \bar{q} (v\overset{{}_\leftarrow}{D})^3 q \rangle
				\end{array}\right)
	=
	\left(\begin{array}{cccc}
		          0 & 1 & 0 & 0\\
		\frac{i}{2} & 1 & 0 & 0\\
		          0 & 0 & 1 & 0\\
		          0 & 0 & 0 & 1
	\end{array}\right)
	\left(\begin{array}{l}
		g^2 \langle \bar q  t^A q \sum_f \bar f \slashed{v} t^A f \rangle\\
		\frac{g}{2} \langle \bar q (v\overset{{}_\leftarrow}{D}) \sigma G q \rangle - m_q^2 \langle \bar q (v\overset{{}_\leftarrow}{D}) q \rangle \\
		\frac{g}{2} \langle \bar q \sigma G (v\overset{{}_\leftarrow}{D}) q \rangle - m_q^2 \langle \bar q (v\overset{{}_\leftarrow}{D}) q \rangle \\
		\frac{1}{v^2} \langle \bar q (v\overset{{}_\leftarrow}{D})^3 q \rangle
	\end{array}\right) \: .
\end{align}
The decomposition of the terms \ref{enum:o2}--\ref{enum:o5} proceeds analogously. The results along with the associated canonical condensates are listed in Appendix~\ref{app:furtherdecompodd}.
\end{enumerate}

The \gls{OPE} relevant combinations of condensates of the odd \gls{OPE} must vanish in vacuum, thus forming the  algebraic vacuum constraints:
\begin{subequations}\label{eq:VacCondCanonOdd}
\label{eq:oddAVLsoft}
\begin{align}
	0 & = \frac{i}{2} g^2 \langle \mathrm{vac} | \bar q t^A q \sum_f \bar f \slashed{v} t^A f | \mathrm{vac} \rangle + g \langle \mathrm{vac} | \bar q (v\overset{{}_\leftarrow}{D}) \sigma G q | \mathrm{vac} \rangle \nonumber\\
	& \quad + \frac{1}{2} g \langle \mathrm{vac} | \bar q \sigma G (v\overset{{}_\leftarrow}{D}) q | \mathrm{vac} \rangle - 3 m_q^2 \langle \mathrm{vac} | \bar q (v\overset{{}_\leftarrow}{D}) q | \mathrm{vac} \rangle
	\label{eq:VacCondCanonOdd1} \: , \\
	0 & = \langle \mathrm{vac} | \bar q (v\overset{{}_\leftarrow}{D})^3 q | \mathrm{vac} \rangle
	\label{eq:VacCondCanonOdd2} \: , \\
	0 & = g \langle \mathrm{vac} | \bar q t^A q \sum_f \bar f \slashed{v} t^A f | \mathrm{vac} \rangle
	\label{eq:VacCondCanonOdd3} \: , \\
	0 & = \langle \mathrm{vac} | \bar q \sigma^{\mu\nu} [(vD) , G_{\mu\nu}] q | \mathrm{vac} \rangle + \langle \mathrm{vac} | \bar q \sigma^{\mu\lambda} [D_\mu , G_{\nu\lambda} v^\nu q | \mathrm{vac} \rangle
	\label{eq:VacCondCanonOdd4} \: , \\
	0 & = \langle \mathrm{vac} | \bar q (v\overset{{}_\leftarrow}{D}) \sigma G q | \mathrm{vac} \rangle .
	\label{eq:VacCondCanonOdd5}
\end{align}
\end{subequations}
Therefore, the condensates in \eqref{eq:VacCondCanonOdd} must vanish individually in vacuum except the condensates in \eqref{eq:VacCondCanonOdd4} which need to cancel each other. Vanishing of the condensates in \eqref{eq:VacCondCanonOdd1} is ensured due to \eqref{eq:VacCondCanonOdd3}, \eqref{eq:VacCondCanonOdd5} and $\langle \mathrm{vac} | \bar q (v\overset{{}_\leftarrow}{D}) q | \mathrm{vac} \rangle = -i m_q \langle \mathrm{vac} | \bar q \slashed{v} q | \mathrm{vac} \rangle$, where the latter condensate is a medium-specific condensate in mass dimension 4 with a zero vacuum value.

\begin{enumerate}[label={\itshape (\roman{*})}, ref={\itshape (\roman{*})}, itemindent=2.5em, leftmargin=0.0em, start=2]
	\label{enum:odd6}
	\item From the next-to-leading order correlator further Gibbs averaged light-quark operators \ref{enum:o6} and \ref{enum:o7} contribute in order $\alpha_\mathrm{s}$
\begin{align}
	&	\langle\!\langle \bar{q} t^A q \bar{q}' \gamma_\mu t^A q' \rangle\!\rangle = \frac{v_\mu}{v^2} \langle\!\langle \bar{q} t^A q \bar{q}' \slashed{v} t^A q' \rangle\!\rangle
	\label{eq:LDecompOddHardFirst} \: , \\
	& \langle\!\langle \bar{q} \gamma_5 \gamma^\mu t^A q \bar{q}' \sigma_{\nu\lambda} t^A q' \rangle\!\rangle = - \frac{1}{6} \varepsilon_{\mu\nu\lambda\alpha} \frac{v^\alpha}{v^2} \langle\!\langle \varepsilon^{\mu'\nu'\lambda'\alpha'} v_{\alpha'} \bar{q} \gamma_5 \gamma_{\mu'} t^A q \bar{q}' \sigma_{\nu'\lambda'} t^A q' \rangle\!\rangle \: ,
	\label{eq:LDecompOddHardLast}
\end{align}
originating from two cut quark lines in the corresponding diagrams.
They are already in canonical form and must vanish in vacuum.
\end{enumerate}

Vacuum constraints of the odd \gls{OPE} of heavy-light meson currents can be deduced from the results presented in this subsection.
Besides Gibbs averaged light-quark operators also heavy-quark operators and such quantities containing both kinds of quark operators enter the \gls{OPE} of heavy-light mesons.
The Gibbs averages \ref{enum:o1}--\ref{enum:o5} are supplemented by similar contributions where $\bar q \overset{{}_\leftarrow}{D}_\mu \cdots q \longrightarrow \bar Q \cdots D_\mu Q$.
The four-quark terms \ref{enum:o6} and \ref{enum:o7} are substituted by similar contributions where one $\bar q \cdots q$ pair is changed for one $\bar Q \cdots Q$ pair.
Be aware that, for two-flavour four-quark terms with two different Dirac structures as in \ref{enum:o6} and \ref{enum:o7}, the necessary substitution yields two structures for each term.

The decompositions of the heavy quark terms and their vacuum constraints originating from the leading order correlator can be obtained by $q \longrightarrow Q$ in \eqref{eq:MedSpecCondOdd1}--\eqref{eq:oddAVLsoft}, where this substitution also refers to subscripts.
The decomposition of the heavy-light four-quark terms originating from the next-to-leading order correlator can be obtained for the exchange of one quark pair $\bar q \cdots q \longrightarrow \bar Q \cdots Q$ in \eqref{eq:LDecompOddHardFirst}--\eqref{eq:LDecompOddHardLast}.
As in the odd \gls{OPE} of light-quark currents all heavy-quark condensates must vanish individually for $\rho \rightarrow 0$.
An exception bears the analog of \eqref{eq:VacCondCanonOdd4}, where both condensates chancel each other.\\

Note that the odd \gls{OPE} and, thus, odd condensates drive the splitting of anti-/particle in the medium.
Such splitting may intuitively be understood in an asymmetric ambient strongly interacting medium.
First, heavy quarks are hardly generated dynamically (at least compared to light quarks) and hardly condensate (directly).
Considering a strongly interacting ambient medium consisting mainly of nuclear matter, i.\,e.\ light quarks (not anti-quarks), a meson consisting of a heavy quark and a light anti-quark is affected differently than its anti-meson, which consists of a heavy anti-quark and a light quark.
Similarly, it seems reasonable to assume that at zero density, meson and anti-meson, e.\,g.\ $D^+$ and $D^-$, are degenerate.
As this splitting is driven by the Lorentz-odd \gls{OPE}, it vanishes at zero baryon density for arbitrary temperatures \cite{Bochkarev:1985ex,Hatsuda:1992bv,Buchheim:2015xka}.

\section{Discussion of factorization and ground state saturation hypothesis}
\label{sec:fac}

The common problem of presently poorly known numerical values of higher mass dimension condensates is often circumvented by a factorization into condensates of lower mass dimension.
Factorization is based on the ground state saturation hypotheses, i.\,e.\ the ground state is assumed to give the dominant contribution to the condensate when a complete set of hadronic states is inserted.
Factorization has been proved correct for an infinite number of colors, but is questionable for QCD \cite{Launer:1983ib,Eletsky:1992xd,Leupold:2005eq,Drukarev:2012av}.
However, since factorization is widely used to get estimates of the numerical values of four-quark condensates we check if the above derived vacuum constraints can be satisfied by factorized four-quark condensates.

Factorization formulae for single-flavour and two-flavour in-medium condensates are provided in \cite{Jin:1992id}.
Two-flavour four-quark condensates containing the generators $t^A$ of the color group give a zero factorization result.
Within the present work this applies to light-quark condensates with the flavour sum $\sum_f$ for $q \neq f$ and to heavy-light four-quark condensates comprising generators.
The nonzero factorization results of light four-quark condensates employed in the decompositions of \ref{enum:e1}--\ref{enum:e5} and \ref{enum:o1}--\ref{enum:o5} read
\begin{subequations}
\begin{align}
	\langle\!\langle \bar q \gamma^\mu t^A q \sum_f \bar f \gamma_\mu t^A f \rangle\!\rangle & =  -\frac{2}{9} \kappa^q_0 (\rho) \left[ 2 \langle\!\langle \bar{q}q \rangle\!\rangle^2 - \langle\!\langle \bar{q}\slashed{v}q \rangle\!\rangle^2/v^2 \right] 
	\label{eq:facCondVac} \: , \\
	\langle\!\langle \bar q \slashed{v} t^A q \sum_f \bar f \slashed{v} t^A f \rangle\!\rangle/v^2 & = -\frac{1}{9} \kappa^q_1 (\rho) \left[ \langle\!\langle \bar{q}q \rangle\!\rangle^2 + \langle\!\langle \bar{q}\slashed{v}q \rangle\!\rangle^2/v^2 \right]
	\label{eq:facCondMed} \: , \\
	\langle\!\langle \bar q t^A q \sum_f \bar f \slashed{v} t^A f \rangle\!\rangle & = -\frac{2}{9} \kappa^q_2 (\rho) \langle\!\langle \bar{q}q \rangle\!\rangle \langle\!\langle \bar{q}\slashed{v}q \rangle\!\rangle  \: ,
	\label{eq:facCondMedOdd}
\end{align}
\end{subequations}
where the $\kappa$'s are fudge factors introduced "by hand" to account for deviations from strict factorization, and they depend on the medium parameter $\rho$ as well as the two-quark condensates do.
For $\kappa = 1$, the factorization equations may be understood as estimates/approximation of four-quark condensates, which become exact in the large $N_\mathrm{c}$ limit.
Further light four-quark condensates with the Dirac structures in \ref{enum:e6}, \ref{enum:e7}, \ref{enum:o6} and \ref{enum:o7} factorize as
\begin{subequations}
\begin{align}
	\langle\!\langle \bar q \gamma^\mu t^A q \bar q \gamma_\mu t^A q \rangle\!\rangle & =  -\frac{2}{9} \kappa^q_0 (\rho) \left[ 2 \langle\!\langle \bar{q}q \rangle\!\rangle^2 - \langle\!\langle \bar{q}\slashed{v}q \rangle\!\rangle^2/v^2 \right] \: , \\
	\langle\!\langle \bar q \slashed{v} t^A q \bar q \slashed{v} t^A q \rangle\!\rangle/v^2 & = -\frac{1}{9} \kappa^q_1 (\rho) \left[ \langle\!\langle \bar{q}q \rangle\!\rangle^2 + \langle\!\langle \bar{q}\slashed{v}q \rangle\!\rangle^2/v^2 \right] \: , \\
	\langle\!\langle \bar q t^A q \bar q \slashed{v} t^A q \rangle\!\rangle & = -\frac{2}{9} \kappa^q_2 (\rho) \langle\!\langle \bar{q}q \rangle\!\rangle \langle\!\langle \bar{q}\slashed{v}q \rangle\!\rangle \: , 
\end{align}
\begin{align}
	\langle\!\langle \bar q \gamma_5 \gamma^\mu t^A q \bar q \gamma_5\gamma_\mu t^A q \rangle\!\rangle & =  \frac{2}{9} \kappa^q_3 (\rho) \left[ 2 \langle\!\langle \bar{q}q \rangle\!\rangle^2 + \langle\!\langle \bar{q}\slashed{v}q \rangle\!\rangle^2/v^2 \right] \: , \\
	\langle\!\langle \bar q \gamma_5 \slashed{v} t^A q \bar q \gamma_5 \slashed{v} t^A q \rangle\!\rangle/v^2 & = \frac{1}{9} \kappa^q_4 (\rho) \left[ \langle\!\langle \bar{q}q \rangle\!\rangle^2 - \langle\!\langle \bar{q}\slashed{v}q \rangle\!\rangle^2/v^2 \right] \: , \\
	\langle\!\langle \varepsilon^{\mu\nu\lambda\alpha} v_\alpha \bar q \gamma_5 \gamma_\mu t^A q \bar q \sigma_{\nu\lambda} t^A q \rangle\!\rangle & = - \frac{16}{3} \kappa^q_5 (\rho) \langle\!\langle \bar{q}q \rangle\!\rangle \langle\!\langle \bar{q}\slashed{v}q \rangle\!\rangle \: , \\
	\langle\!\langle \bar q \gamma^\mu q \bar q \gamma_\mu q \rangle\!\rangle & =  -\frac{1}{6} \kappa^q_6 (\rho) \left[ 2 \langle\!\langle \bar{q}q \rangle\!\rangle^2 - 7 \langle\!\langle \bar{q}\slashed{v}q \rangle\!\rangle^2/v^2 \right] \: , \\
	\langle\!\langle \bar q \slashed{v} q \bar q \slashed{v} q \rangle\!\rangle/v^2 & = -\frac{1}{12} \kappa^q_7 (\rho) \left[ \langle\!\langle \bar{q}q \rangle\!\rangle^2 - 11 \langle\!\langle \bar{q}\slashed{v}q \rangle\!\rangle^2/v^2 \right] \: , \\
	\langle\!\langle \bar q q \bar q \slashed{v} q \rangle\!\rangle & = \frac{11}{12} \kappa^q_8 (\rho) \langle\!\langle \bar{q}q \rangle\!\rangle \langle\!\langle \bar{q}\slashed{v}q \rangle\!\rangle \: , \\
	\langle\!\langle \bar q \gamma_5 \gamma^\mu q \bar q \gamma_5\gamma_\mu q \rangle\!\rangle & =  \frac{1}{6} \kappa^q_9 (\rho) \left[ 2 \langle\!\langle \bar{q}q \rangle\!\rangle^2 + \langle\!\langle \bar{q}\slashed{v}q \rangle\!\rangle^2/v^2 \right] \: , \\
	\langle\!\langle \bar q \gamma_5 \slashed{v} q \bar q \gamma_5 \slashed{v} q \rangle\!\rangle/v^2 & = \frac{1}{12} \kappa^q_{10} (\rho) \left[ \langle\!\langle \bar{q}q \rangle\!\rangle^2 - \langle\!\langle \bar{q}\slashed{v}q \rangle\!\rangle^2/v^2 \right] \: , \\
	\langle\!\langle \varepsilon^{\mu\nu\lambda\alpha} v_\alpha \bar q \gamma_5 \gamma_\mu q \bar q \sigma_{\nu\lambda} q \rangle\!\rangle & = -4 \kappa^q_{11} (\rho) \langle\!\langle \bar{q}q \rangle\!\rangle \langle\!\langle \bar{q}\slashed{v}q \rangle\!\rangle \: .
\end{align}
\end{subequations}
For condensates composed of heavy and light quark operators with the Dirac structures in \ref{enum:e6}, \ref{enum:e7}, \ref{enum:o6} and \ref{enum:o7} containing the unit elements $T^0 = 1$ as color structures one obtains \cite{Jin:1992id}:
\begin{subequations}
\label{eq:facCondUnitElem}
\begin{align}
	\langle\!\langle \bar q \gamma_\mu q \bar Q \gamma^\mu Q \rangle\!\rangle & = \kappa^Q_0(\rho) \left[ \langle\!\langle \bar q q \rangle\!\rangle \langle\!\langle \bar Q Q \rangle\!\rangle + \langle\!\langle \bar q \slashed{v} q \rangle\!\rangle \langle\!\langle \bar Q\slashed{v} Q \rangle\!\rangle / v^2 \right] \: , \\
	\langle\!\langle \bar q \slashed{v} q \bar Q \slashed{v} Q \rangle\!\rangle / v^2 & = \frac{1}{4} \kappa^Q_1(\rho) \left[ \langle\!\langle \bar q q \rangle\!\rangle \langle\!\langle \bar Q Q \rangle\!\rangle + 4 \langle\!\langle \bar q \slashed{v} q \rangle\!\rangle \langle\!\langle \bar Q\slashed{v} Q \rangle\!\rangle / v^2 \right] \: , \\
	\langle\!\langle \bar q q \bar Q \slashed{v} Q \rangle\!\rangle & = \kappa^Q_2(\rho) \langle\!\langle \bar q q \rangle\!\rangle \langle\!\langle \bar Q \slashed{v} Q \rangle\!\rangle \: , \\
	\langle\!\langle \bar q \slashed{v} q \bar Q Q \rangle\!\rangle & = \kappa^Q_3(\rho) \langle\!\langle \bar q \slashed{v} q \rangle\!\rangle \langle\!\langle \bar Q Q \rangle\!\rangle \: , \\
	\langle\!\langle \bar q \gamma_5 \gamma_\mu q \bar Q \gamma_5 \gamma^\mu Q \rangle\!\rangle & = - \kappa^Q_4(\rho) \langle\!\langle \bar q q \rangle\!\rangle \langle\!\langle \bar Q Q \rangle\!\rangle \: , \\
	\langle\!\langle \bar q \gamma_5 \slashed{v} q \bar Q \gamma_5 \slashed{v} Q \rangle\!\rangle / v^2 & = - \frac{1}{4} \kappa^Q_5(\rho) \langle\!\langle \bar q q \rangle\!\rangle \langle\!\langle \bar Q Q \rangle\!\rangle \: , \\
	\langle\!\langle \varepsilon^{\mu\nu\lambda\alpha} v_\alpha \bar q \gamma_5 \gamma_\mu q \bar Q \sigma_{\nu\lambda} Q \rangle\!\rangle & = 0 \: , \\
	\langle\!\langle \varepsilon^{\mu\nu\lambda\alpha} v_\alpha \bar q \sigma_{\mu\nu} q \bar Q \gamma_5 \gamma_\lambda Q \rangle\!\rangle & = 0 \: .
\end{align}
\end{subequations}
The same relations hold true for $Q \longrightarrow q'$ providing the factorization formulae for light two-flavour condensates.

In vacuum, where $\langle\!\langle \cdots \rangle\!\rangle$ reduces to $\langle \mathrm{vac} | \cdots | \mathrm{vac} \rangle$, the medium-specific two-quark condensate $\langle\!\langle \bar{q}\slashed{v}q \rangle\!\rangle$ cancels, thus further simplifying above relations and yielding zero factorization results.
The vacuum constraint \eqref{eq:evenAVLsoft} from the even \gls{OPE}, where light four-quark condensates are factorized, provide the following relations 
\begin{subequations}
\label{eq:even}
\begin{align}
	& g \langle \mathrm{vac} | \bar{q} (vi\overset{{}_\leftarrow}{D}) \sigma G \slashed{v} q | \mathrm{vac} \rangle / v^2 =
	\frac{1}{4} \bigg( 4 m_q^3 \langle \mathrm{vac} | \bar q q | \mathrm{vac} \rangle - g m_q \langle \mathrm{vac} | \bar q \sigma G q | \mathrm{vac} \rangle \bigg) \: , \\
	& g \langle \mathrm{vac} | \bar{q} (vi\overset{{}_\leftarrow}{D}) \gamma^\mu G_{\nu\mu} v^\nu q | \mathrm{vac} \rangle / v^2 =
	- \frac{1}{54} \bigg( 2 \kappa^q_0(0) - \kappa^q_1(0) \bigg) g^2 \langle \mathrm{vac} | \bar q q | \mathrm{vac} \rangle^2 \: , \\
	& g \langle \mathrm{vac} | \bar{q} \gamma^\mu [(vD),G_{\mu\nu}] v^\nu q | \mathrm{vac} \rangle / v^2 =
	- \frac{1}{9} \bigg( 2 \kappa^q_0(0) - \kappa^q_1(0) \bigg) g^2 \langle \mathrm{vac} | \bar q q | \mathrm{vac} \rangle^2 \: , 
	\label{eq:facexample}
\end{align}
\begin{align}
	& g \langle \mathrm{vac} | \bar{q} \slashed{v} \sigma^{\nu\lambda} [(viD),G_{\nu\lambda}] q | \mathrm{vac} \rangle / v^2 =
	- \frac{2}{9} \bigg( \kappa^q_0(0) - \kappa^q_1(0) \bigg) g^2 \langle \mathrm{vac} | \bar q q | \mathrm{vac} \rangle^2 \: , 
	\label{eq:zeroavlfac}\\
	& g \langle \mathrm{vac} | \varepsilon_{\mu\nu\lambda\tau} v^\tau \bar{q} i \overset{{}_\leftarrow}{D}{}^\mu \gamma_5 \slashed{v} G^{\nu\lambda} q | \mathrm{vac} \rangle / v^2 = 
	\frac{1}{12} \bigg( 6 m_q^3 \langle \mathrm{vac} | \bar q q | \mathrm{vac} \rangle - 8 g m_q \langle \mathrm{vac} | \bar q \sigma G q | \mathrm{vac} \rangle \nonumber\\
	& \quad + \frac{4}{9} \left( 7 \kappa^q_0(0) - 2 \kappa^q_1(0) \right) g^2 \langle \mathrm{vac} | \bar q q | \mathrm{vac} \rangle^2 - 24 y_q \bigg) \: , \\
	& \langle \mathrm{vac} | \bar{q} (vi\overset{{}_\leftarrow}{D})^3 \slashed{v} q | \mathrm{vac} \rangle / v^4 =
	\frac{1}{48} \bigg( -6 m_q^3 \langle \mathrm{vac} | \bar q q | \mathrm{vac} \rangle + 3 g m_q \langle \mathrm{vac} | \bar q \sigma G q | \mathrm{vac} \rangle \nonumber\\
	& \quad - \frac{4}{9} \kappa^q_0(0) g^2 \langle \mathrm{vac} | \bar q q | \mathrm{vac} \rangle^2 \bigg)
\end{align}
\end{subequations}
supplemented by six relations contained in \eqref{eq:SolVacConNLOCorr} (originating from next-to-leading order correlator)
\begin{subequations}
\label{eq:evenNLOcorr}
\begin{align}
	\kappa^q_1(0) &= \kappa^q_0(0)
	\label{eq:VacMedLight4qRel} \: , \\
	\kappa^q_4 (0) &= \kappa^q_3 (0) \: , \\
	\kappa^q_7 (0) &= \kappa^q_6 (0) \: , \\
	\kappa^q_{10} (0) &= \kappa^q_9 (0) \: , \\[-0.4ex]
	\kappa^{q'}_1(0) &= \kappa^{q'}_0(0) \: , \\[-0.4ex]
	\kappa^{q'}_5(0) &= \kappa^{q'}_4(0) \: ,
\end{align}
\end{subequations}
relating the $\kappa$ parameters for $\rho=0$ and simplifying the factorized algebraic vacuum constraints \eqref{eq:even}, where \eqref{eq:zeroavlfac} gives a vanishing \gls{VEV} as can be recognised already in Eqs.~\eqref{eq:zeroavl} and \eqref{eq:SolVacConNLOCorr}.
The odd \gls{OPE} vacuum constraints are trivial as the contained four-quark condensates vanish individually in the vacuum.

In heavy-light current \glspl{OPE}, the vacuum constraints in factorized form are composed of the relations \eqref{eq:even} as well as these relations with  $\kappa^q_{0,1}=0$ and $q \longrightarrow Q$.
The vacuum constraints \eqref{eq:evenNLOcorr} are substituted by
\begin{subequations}
\label{eq:evenNLOcorrhl}
\begin{align}
	& \kappa^Q_1(0) = \kappa^Q_0(0) \: , \label{eq:facKappaRel}\\
	& \kappa^Q_5(0) = \kappa^Q_4(0) \: .
\end{align}
\end{subequations}

As exhibited in Eqs.~\eqref{eq:even}--\eqref{eq:evenNLOcorrhl} factorization of four-quark condensates does not violate the algebraic vacuum constraints.
Instead, it relates the $\kappa$ factorization parameters at $\rho=0$ and assigns numerical vacuum values to the medium dimension 6 condensates which were previously unknown.
Relating it to the known vacuum values of the chiral and the mixed quark-gluon condensate allows for their tentative numerical estimate in vacuum.

In Fig.~\ref{fig:fac} we employ the factorization formulae \eqref{eq:facCondVac}, \eqref{eq:facCondMed} and \eqref{eq:facexample} to illustrate the numerical estimates of the vacuum-specific and medium-specific condensates entering the in-medium decomposition of the Gibbs average \ref{enum:e2}.
After transformation to canonical condensates utilizing Eq.~\eqref{eq:enum2CanonCondTrafo} the vacuum-specific condensate contained in Eq.~\eqref{eq:VacSpecCond2} reads%
\begin{align}
	\langle\!\langle g \bar q \gamma_\mu t^A q \sum_f \bar f \gamma^\mu t^A f \rangle\!\rangle\nonumber
\end{align}
while the medium-specific condensate contained in Eq.~\eqref{eq:MedSpecCond2} is
%
%
\begin{align}
	& \langle\!\langle g \bar q \gamma_\mu t^A q \sum_f \bar f \gamma^\mu t^A f - \frac{2}{v^2}\Big( \bar q \gamma^\alpha[(vD),G_{\alpha\beta}]v^\beta q + g \bar q \slashed{v} t^A q \sum_f \bar f \slashed{v} t^A f \Big) \rangle\!\rangle \nonumber
\end{align}
\begin{align}\label{eq:examplMedSpezCond}
	& \quad \equiv \langle\!\langle g \bar q \gamma_\mu t^A q \sum_f \bar f \gamma^\mu t^A f \rangle\!\rangle - \frac{2}{v^2}\Big( \langle\!\langle \bar q \gamma^\alpha[(vD),G_{\alpha\beta}]v^\beta q \rangle\!\rangle + \langle\!\langle g \bar q \slashed{v} t^A q \sum_f \bar f \slashed{v} t^A f \rangle\!\rangle \Big) \: .
\end{align}
Equations~\eqref{eq:facCondVac} and \eqref{eq:facCondMed} are in-medium relations suitable to deduce the medium dependencies of the four-quark condensates from the medium behaviors of the two-quark condensates.
In the rest frame, where $v_{\mu} = (1,\vec{0})$, one obtains $\langle\!\langle \bar{q}q \rangle\!\rangle = \langle \mathrm{vac} | \bar q q | \mathrm{vac} \rangle \big( 1 - \frac{\sigma_N}{m_\pi^2f_\pi^2} \rho_N \big)$ and $\langle\!\langle \bar{q}\slashed{v}q \rangle\!\rangle = \langle\!\langle q^\dagger q \rangle\!\rangle = \frac{3}{2}\rho_N$ in leading order of the net nucleon density $\rho_N$ \cite{Jin:1992id}.
The quantities $f_\pi$ and $m_\pi$ denote the pion decay constant and pion mass, respectively, while $\sigma_N$ is the nucleon sigma term.
The four-quark terms acquire the following density dependencies:
\begin{align}
	\langle\!\langle g \bar q \gamma_\mu t^A q \sum_f \bar f \gamma^\mu t^A f \rangle\!\rangle & = -g \frac{4}{9} \kappa^q_0(\rho_N) \left[ \langle \mathrm{vac} | \bar q q | \mathrm{vac} \rangle^2 \left( 1 - \frac{\sigma_N}{m_\pi^2f_\pi^2} \rho_N \right)^2 - \frac{9}{8} \rho_N^2 \right]\: , \\
	\langle\!\langle g \bar q \slashed{v} t^A q \sum_f \bar f \slashed{v} t^A f \rangle\!\rangle/v^2 & = -g \frac{1}{9} \kappa^q_1(\rho_N) \left[ \langle \mathrm{vac} | \bar q q | \mathrm{vac} \rangle^2 \left( 1 - \frac{\sigma_N}{m_\pi^2f_\pi^2} \rho_N \right)^2 + \frac{9}{4} \rho_N^2 \right] \: .
\end{align}
However, the \gls{EV} of the third constituent of the medium-specific condensate \eqref{eq:examplMedSpezCond}, the quark-gluon condensate $\langle\!\langle \bar q \gamma^\alpha[(vD),G_{\alpha\beta}]v^\beta q \rangle\!\rangle/v^2$, is only known in vacuum from the algebraic vacuum limit~\eqref{eq:facexample}.
We introduce "by hand" a linear density dependence of this medium condensate modifying its vacuum value $\langle\!\langle \bar q \gamma^\alpha[(vD),G_{\alpha\beta}]v^\beta q \rangle\!\rangle/v^2 = \langle\ \mathrm{vac} | \bar q \gamma^\alpha[(vD),G_{\alpha\beta}]v^\beta q | \mathrm{vac} \rangle/v^2 \; (1 + m\,\rho_N)$ with $m \in [-2,2]$ in order to exemplify a potential medium dependence of this particular condensate.
The range of $m$ is in approximate numerical agreement with the linear density dependencies of the two-quark condensates.
Accordingly, the quark-gluon condensate in (dense) nuclear matter reads
\begin{align}\label{eq:qgDensDep}
	\langle\!\langle \bar q \gamma^\alpha[(vD),G_{\alpha\beta}]v^\beta q \rangle\!\rangle/v^2 = - \frac{1}{9} \bigg( 2 \kappa^q_0(0) - \kappa^q_1(0) \bigg) g \langle \mathrm{vac} | \bar q q | \mathrm{vac} \rangle^2 \; (1 + m\,\rho_N) \: .
\end{align}
For the $\kappa$ parameters we choose $\kappa^q_i(\rho_N) = \kappa^q_i(0) = 1$ for $i=0$ and 1 in accordance with relation \eqref{eq:facKappaRel}.
The vacuum expectation value of the chiral condensate $\langle \mathrm{vac} | \bar q q | \mathrm{vac} \rangle = (-0.245\; \mathrm{GeV})^3$ is employed as well as $g = \sqrt{4\pi\alpha_\mathrm{s}}$ with $\alpha_\mathrm{s} = 0.5$.
The above derived formulae \eqref{eq:examplMedSpezCond}--\eqref{eq:qgDensDep} are utilized to illustrate the important features of vacuum-specific and medium-specific condensates in Fig.~\ref{fig:fac}.

\section{Discussion of heavy-quark expansion}
\label{sec:hqe}

Besides light-quark condensates also heavy-quark condensates enter open-flavor \glspl{QSR} containing a light and a heavy quark \cite{Zschocke:2011aa}.
Heavy quarks do hardly condense since they are too heavy to be produced from the QCD vacuum \cite{Reinders:1984sr}.
However, they can interact with the ground state via soft gluons.
Accordingly, heavy-quark condensates can be expanded into a series of gluonic condensates weighted by powers of the inverse heavy-quark mass $m_Q$.
This procedure is dubbed \gls{HQE}.
It is well established for vacuum heavy-quark condensates \cite{Generalis:1983hb,Broadhurst:1985js,Bagan:1984zt,Bagan:1985zp}, such as the heavy two-quark condensate $\langle \mathrm{vac} | \bar Q Q | \mathrm{vac} \rangle$, and has been extended to cover medium four-quark condensates consisting of a light and a heavy-quark pair \cite{Buchheim:2014uda}.
In order to check the compatibility of the \gls{HQE} with the full set of algebraic vacuum limits, inter alia containing heavy-quark condensates with covariant derivatives, the approach presented in Ref.~\cite{Grozin:1994hd} has to be extended to medium condensates.
\begin{figure}%
\begin{minipage}[b]{0.486\columnwidth}
	\captionsetup{aboveskip=5.7ex}
	\begin{subfigure}{0.59\columnwidth}
	\captionsetup{aboveskip=2ex}
	\includegraphics[width=1.\columnwidth]{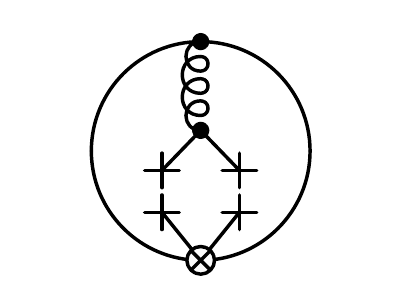}%
	\caption{LO}
	\label{fig:lo}
	\end{subfigure}
	\hspace{-0.21\columnwidth}
	\begin{subfigure}{0.59\columnwidth}
	\captionsetup{aboveskip=2ex}
	\includegraphics[width=1.\columnwidth]{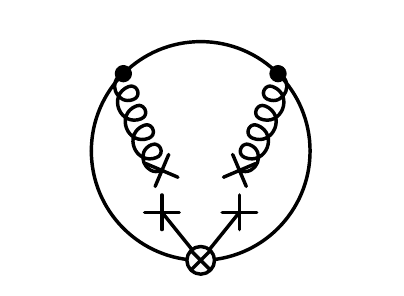}%
	\caption{NLO}
	\label{fig:nlo}
	\end{subfigure}
\caption{Diagrammatic representa\-tion of the leading order (LO) and next-to-leading order (NLO) \gls{HQE} contributions to the heavy-light four-quark condensate.}%
\label{fig:hqediagrams}%
\end{minipage}
\hfill
\begin{minipage}[b]{0.514\columnwidth}
\includegraphics[width=1.0\columnwidth]{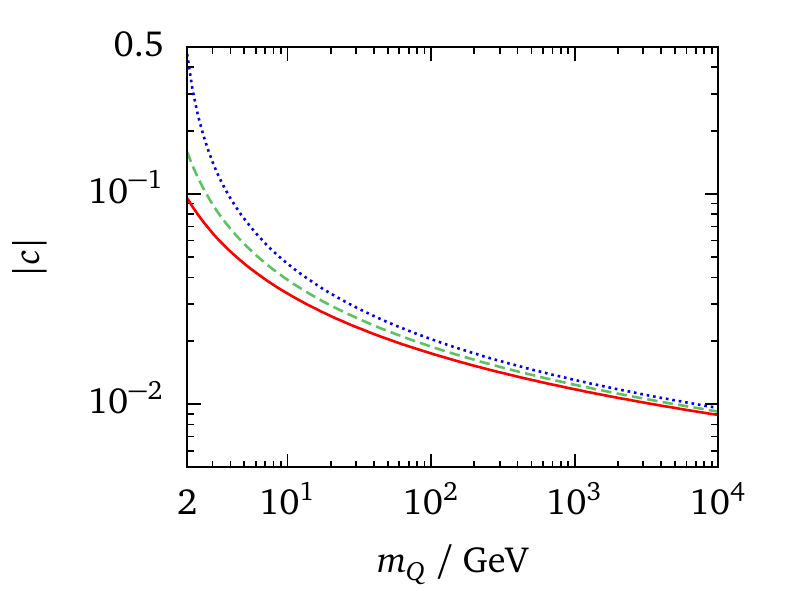}%
\caption{The absolute value of $c$ defined in \eqref{eq:c} as a function of $m_Q$. The red solid, green dashed and blue dotted curves correspond to $\mu^2=0.5$, 1 and 2 $\mathrm{GeV}^2$, respectively.}%
\label{fig:hqe}%
\end{minipage}
\end{figure}

We choose the algebraic vacuum constraint \eqref{eq:SolVacConNLOCorr}, with $\Gamma'=1$, $T^A=t^A$ and one $\bar q \ldots q$ pair substituted by a $\bar Q \ldots Q$ pair, entering the heavy-light quark current \gls{OPE}
\begin{align}\label{eq:hlvaclim}
	\langle \mathrm{vac} | \bar q \slashed{v} t^A q \bar Q \slashed{v} t^A Q | \mathrm{vac} \rangle / v^2 = \frac{1}{4} \langle \mathrm{vac} | \bar q \gamma_\mu t^A q \bar Q \gamma^\mu t^A Q | \mathrm{vac} \rangle
\end{align}
as the first algebraic vacuum limit deduced by the above formalism to be checked for compatibility with \gls{HQE}.
Utilizing the leading order \gls{HQE} formulae for both sides of Eq.~\eqref{eq:hlvaclim} separately, i.\,e.\
\begin{align}
	\langle\!\langle \bar q \slashed{v} t^A q \bar Q \slashed{v} t^A Q \rangle\!\rangle / v^2 & = - \frac{2}{3} \frac{g^2}{(4\pi)^2} \bigg( \ln\frac{\mu^2}{m_Q^2} + \frac{1}{3} \bigg) \langle\!\langle \bar q \slashed{v} t^A q \sum_f \bar f \slashed{v} t^A f \rangle\!\rangle / v^2 + \cdots\: ,\\
	\langle\!\langle \bar q \gamma_\mu t^A q \bar Q \gamma^\mu t^A Q \rangle\!\rangle & = - \frac{2}{3} \frac{g^2}{(4\pi)^2} \bigg( \ln\frac{\mu^2}{m_Q^2} + \frac{1}{2} \bigg) \langle\!\langle \bar q \gamma_\mu t^A q \sum_f \bar f \gamma^\mu t^A f \rangle\!\rangle + \cdots
\end{align}
calculated in the $\overline{\mathrm{MS}}$ regularisation scheme\footnote{Schematically, the evaluation of the 1-point function
$\langle\!\langle \bar q q \bar Q Q \rangle\!\rangle
= C_\mathrm{LO} \langle\!\langle \bar q q \sum_ f\bar f f \rangle\!\rangle
+ C_\mathrm{NLO} \langle\!\langle \bar q  q G^2 \rangle\!\rangle 
+ \cdots$
is performed, where leading order (LO) and next-to-leading order (NLO) \gls{HQE} contributions, requiring the evaluation of the diagrams depicted in Figs.~\ref{fig:hqediagrams}, are displayed explicitly and higher orders are denoted by dots \cite{Buchheim:2014uda}. The quantity $\mu$ is the renormalisation scale usually chosen to be of the order of $1\;\mathrm{GeV}$ in hadron physics \cite{Reinders:1984sr}.} one obtains from \eqref{eq:hlvaclim} upon $\lim\limits_{\rho \rightarrow 0} \langle\!\langle \cdots \rangle\!\rangle = \langle \mathrm{vac} | \cdots | \mathrm{vac} \rangle$
\begin{align}
	\langle \mathrm{vac} | \bar q \slashed{v} t^A q \sum_f \bar f \slashed{v} t^A f | \mathrm{vac} \rangle / v^2 = \frac{1 + c}{4} \langle \mathrm{vac} | \bar q \gamma_\mu t^A q \sum_f \bar f \gamma^\mu t^A f | \mathrm{vac} \rangle
\end{align}
with
\begin{align}\label{eq:c}
	c = \frac{1}{2} \bigg( 3 \ln\frac{\mu^2}{m_Q^2} + 1 \bigg)^{-1} \: .
\end{align}
However, the algebraic vacuum constraint \eqref{eq:SolVacConNLOCorr}, with $\Gamma'=1$ and $T^A=t^A$, entering the light-quark current \gls{OPE} is satisfied solely for $c=0$. 
Instead, $c=c(m_Q)$ runs only logarithmically to zero for increasing heavy-quark masses $m_Q$ as exhibited in Fig.~\ref{fig:hqe}.
Only in the limit of an infinite heavy-quark mass the algebraic vacuum limits deduced from medium-specific condensates are compatible with leading order \gls{HQE} for this specific example, albeit showing strong deviations from $c=0$ for experimentally constrained values caused by the logarithmic \gls{HQE} terms.

\section{Comments on chirally odd condensates}
\label{sec:chiral}

Chiral transformations act as flavour rotations on the left and right handed quark spinors separately building the $\mathrm{SU}(N_\mathrm{f})_\mathrm{L} \times \mathrm{SU}(N_\mathrm{f})_\mathrm{R}$ symmetry group, where $N_\mathrm{f}$ is the number of quark flavours.
QCD condensates which are not invariant under chiral transformations, such as the chiral condensate, can be transformed into their negatives by a suitable transformation.
Thus, they are dubbed "chirally odd" and, in the spirit of \glspl{QSR}, they quantify the difference of chiral partner spectra, i.\,e.\ they attain a zero value if chiral symmetry is restored.
Accordingly, condensates which are invariant under chiral transformation are referred to as "chirally even".
The notions of chiral symmetry are not restricted to light hadrons but can be applied to open-flavour mesons as well \cite{Hilger:2011cq,Buchheim:2015xka}.

Besides chirally odd four-quark condensates, e.\,g.\ discussed in Ref.~\cite{Hilger:2010cn}, also chirally even four-quark condensates exist.
Factorization of these condensates into the square of the chiral condensate, which is genuinely chirally odd, distorts their behavior under chiral transformations.
Apart from lacking accuracy, factorization is also disputable with respect to chiral symmetry.

All chirally odd condensates of a certain mass dimension $\dim_m$ appear either in the even or in the odd \gls{OPE} of the leading order correlator, i.\,e.\ they have either an even or odd Lorentz tensor rank $n$. 
The four-quark condensate given e.\,g.\ in \eqref{eq:chioddcond} is a particular example of a condensate which is characterized by its non-trivial chiral transformation properties, similar to the famous chiral condensate $\langle\!\langle \bar q q \rangle\!\rangle$, as an order parameter of \gls{DCSB}.
However, as it is a Lorentz-odd condensate, it must be zero at any temperature for zero net density if particle and anti-particle are to remain degenerate (see Sec.\,\ref{subsec:oddOPE}).
This exemplifies why the vanishing of a chirally odd condensate, such as the chiral condensate, is not a sufficient condition for chiral symmetry restoration if only the non-trivial chiral transformation properties of such a condensate are taken into account \cite{Hilger:2015zva}.

This can be deduced from chiral partner (c.p.) \glspl{OPE} where chirally even condensates cancel.
In the \glspl{EV} of Eqs.~(2.6b) and (2.6d) in reference \cite{Hilger:2011cq} only elements of the Clifford algebra $\Gamma$ with an even number of Lorentz indices enter:
\begin{align}
	\label{eq:chioddcontrib}
	\Pi^{(2)}_\mathrm{c.p.}
	\propto
	\sum_{k,l,m} \sum_\Gamma^{\{1,\sigma_{\alpha<\beta},\gamma_5\}} \langle \bar q \overset{{}_\leftarrow}{D}_{\alpha_1} \cdots \overset{{}_\leftarrow}{D}_{\alpha_k} \Gamma S (D_{\mu_1},\ldots,D_{\mu_l};G_{\nu_1\lambda_1},\ldots,G_{\nu_m\lambda_m}) q \rangle \: ,
\end{align}
where the mass dimension $\dim_m$ of the \gls{EV} is given by $\dim_m=3+k+l+2m$.\footnote{A more detailed mass formula which is based on the order of the quark field expansion and the expansion of the perturbative quark propagator can be found in \cite{UweHilger:2012uua}. Note that Eq.~\eqref{eq:chioddcontrib} holds true solely for quark condensates from the leading order correlator.}
In the above relation $S$ denotes a term of the expansion of the perturbative quark propagator in a weak gluonic background field specified by $l$ and $m$.
It is merely a function of the field strength tensor and derivatives thereof.
Thus, the Dirac matrices $\Gamma$ do not alter the spin parity of the \gls{EV}.
Since the number of Lorentz indices and the order of mass dimension are increased by the same integer if $G_{\mu\nu}$ and $D_\mu$ are added, one can stick with the mass dimension of the individual operators contributing to the \gls{EV} to determine the rank of this Lorentz tensor.
Thus, the Lorentz tensor rank of the \gls{EV} is $n = \dim_m-3$, where 3 is the mass dimension of the two quark operators.
From relation~\eqref{eq:chioddcontrib} we, therefore, infer for 
that all chirally odd condensates of this mass dimension enter the odd (even) \gls{OPE}, due to their odd (even) tensor rank $n$.

\section{Summary}
\label{sec:summary}

\glsresetall

\glspl{QSR} rely on the \gls{OPE} of current-current correlators.
The \gls{OPE} allows to separate a perturbatively accessible short-range behavior (encoded in Wilson coefficients) and a genuinely non-perturbative long-range part (encoded in a set of QCD operators).
\Acrlongpl{VEV} or Gibbs averages of the operators lead to condensates (see \cite{Brodsky:2009zd,Brodsky:2010xf,Brodsky:2012ku,Chang:2013epa,Cloet:2013jya} for a discussion of the meaning of condensates).
The emerging operators (or, subsequently, condensates) are, in general, Lorentz tensors.
It is the aim of the present paper to provide a formalized manner to handle these tensor structures with the goal to ensure a clean transition from an in-medium situation to the vacuum.
In doing so we define medium-specific scalar operators and condensates which, by definition, vanish in vacuum.
These must be distinguished from the medium dependence of a condensate.
A vacuum-specific condensate, in contrast, occurs already in vacuum and can obey a medium dependence too.

While the explication of our formalism looks quite cumbersome, it appears as fairly practical when discussing such issues as factorization and heavy-quark expansion.
The need to account for higher-rank tensors comes from the inclusion of higher mass dimensions, i.\,e.\ shifting the truncation of the \glspl{QSR} to higher orders.
Besides the question of the impact of higher-order terms, the determination of a Borel window requires the comparison of low-order with high-order terms, for instance, in applications of Borel-transformed \glspl{QSR}.
The recent investigation of the four-quark condensates in \gls{QSR} applications to open-charm mesons \cite{Buchheim:2014rpa} is at the very motivation to make our formalism explicit for these contributions.
This evidences that beyond the intricate calculation of Wilson coefficients also the proper handling of condensates with non-trivial Lorentz tensor structures is indispensable, in particular, when considering the relation of in-medium \glspl{QSR} and such ones in vacuum.

Not only the planned facilities at NICA, FAIR and J-PARC are going to address charm degrees of freedom in a baryon-dense medium, but the running collider experiments at LHC and RHIC are delivering at present a wealth of data on charm and bottom degrees of freedom in a high-temperature environment at very small net baryon density.
The firm application of \glspl{QSR} on these quite different experimental conditions and the relation to observables deserve much more dedicated investigations on the theory side.

\section*{Acknowledgments}
The authors gratefully acknowledge enlightening discussions with W.\ Weise, S.\ Leupold, S.\ Brodsky, S.\ H.\ Lee, K.\ Morita, R.\ Rapp, and S.\ Zschocke. 
The work is supported by BMBF grant 05P12CRGHE and by the Austrian Science Fund (FWF) under project number P25121-N27.

\appendix

\section{Equivalence of two approaches to the medium-specific decomposition}
\label{app:equiv}

One may find the medium-specific decomposition either by subtraction of the algebraic vacuum decomposition \eqref{eq:genprojalg} from the complete in-medium decomposition \eqref{eq:defplain} using Eq.~\eqref{eq:defmed} or by performing the projection procedure for the medium-specific decomposition structures in \eqref{eq:medprojvec} which are orthogonal to the vacuum decomposition structures. Following the former approach we show that utilizing the blockwise inversion of the total projection matrix
\begin{align}
	\label{eq:totprojmatrix}
	L_\rho \equiv \left( \vec l^{(\rho)} \circ \vec l^{(\rho)} \right)
					= \left(\begin{array}{cc}
										\left( \vec l^{(0)} \circ \vec l^{(0)} \right)	&	\left( \vec l^{(0)} \circ \vec l^{(\rho_1)} \right)\\
										\left( \vec l^{(\rho_1)} \circ \vec l^{(0)} \right)	&	\left( \vec l^{(\rho_1)} \circ \vec l^{(\rho_1)} \right)
									\end{array}\right) 
					= \left(\begin{array}{cc}
										L_0	&	L_{0,\rho_1}\\
										L_{\rho_1,0}	&	L_{\rho_1}
									\end{array}\right) 
\end{align}
yields the same medium-specific decomposition structures $\vec l^{(1)}_\mu$ as the latter approach based on orthogonality, i.\,e.\ the equivalence of both approaches to the medium-specific decomposition. We rearrange Eq.~\eqref{eq:defmed}:
\begin{align}
	\vec l^{(1)}_\mu \cdot \vec a_1
	=  \vec l^{(\rho)}_\mu \cdot \vec a_{\rho} - \vec l^{(0)}_\mu \cdot \vec a_0 
	= \vec l^{(0)}_\mu \cdot \left( \vec a_{\rho_0} - \vec a_0 \right) + \vec l^{(\rho_1)}_\mu \cdot \vec a_{\rho_1} \: .
\end{align}
Using the projections 
\begin{align}
	\vec a_0 & = L_0^{-1} \vec l^{(0)}_\nu \langle O^\nu \rangle
	\label{eq:projavac}\\
	\vec a_\rho & = L^{-1}_\rho \vec l^{(\rho)}_\nu \langle O^\nu \rangle
	\label{eq:projA}
\end{align}
with the sub-matrix definitions\footnote{Note that \(\left( L^{-1} \right)_{i} \neq L_i^{-1}\) for \( i \in \{0;0,\rho_1;\rho_1,0;\rho_1\} \).} of the inverse total projection matrix
\begin{align}
	\label{eq:invtotprojmatrix}
	L_\rho^{-1} = 	\left(\begin{array}{cc}
										\left( L^{-1} \right)_0	&	\left( L^{-1} \right)_{0,\rho_1}\\
										\left( L^{-1} \right)_{\rho_1,0}	&	\left( L^{-1} \right)_{\rho_1}
									\end{array}\right)
\end{align}
further decomposed to 
\begin{align}
	\vec a_{\rho_0} = \left( \left( L^{-1} \right)_0 , \left( L^{-1} \right)_{0,\rho_1} \right) \vec l^{(\rho)}_\nu \langle O^\nu \rangle
	\quad \text{and} \quad
	\vec a_{\rho_1} = \left( \left( L^{-1} \right)_{\rho_1,0} , \left( L^{-1} \right)_{\rho_1} \right) \vec l^{(\rho)}_\nu  \langle O^\nu \rangle
	\label{eq:proja(rho)decomp}
\end{align}
yields
\begin{subequations}
\begin{align}
	\vec l^{(1)}_\mu \cdot \vec a_1
	& =  \vec l^{(0)}_\mu \cdot \left[ \left( \left( L^{-1} \right)_0 , \left( L^{-1} \right)_{0,\rho_1}  \right) \vec l^{(\rho)}_\nu \langle O^\nu \rangle - L_0^{-1} \vec l^{(0)}_\nu \langle O^\nu \rangle \right] + \vec l^{(\rho_1)}_\mu \cdot \vec a_{\rho_1}\\
	& = \vec l^{(0)}_\mu \cdot \left[ \left( L^{-1} \right)_0 - L_0^{-1} \right] \vec l^{(0)}_\nu \langle O^\nu \rangle + \vec l^{(0)}_\mu \cdot \left( L^{-1} \right)_{0,\rho_1} \vec l^{(\rho_1)}_\nu \langle O^\nu \rangle + \vec l^{(\rho_1)}_\mu \cdot \vec a_{\rho_1} \: .
	\label{eq:Pn-Pnvac}
\end{align}
\end{subequations}
The formula for blockwise inversion of matrices with two square sub-matrices \(A\) and \(D\)
\begin{align}
	\left[
	\begin{array}{cc}
		A & B\\
		C & D
	\end{array}
	\right]^{-1}
	& =
	\left[
	\begin{array}{cc}
		A^{-1} + A^{-1}BSCA^{-1} & - A^{-1}BS\\
		- SCA^{-1} & S
	\end{array}
	\right]
\end{align}
with
\begin{align}
	S & = \left( D - CA^{-1}B \right)^{-1}
\end{align}
relates the sub-matrices of the inverted matrix. We deploy this formula to the total projection matrix \eqref{eq:totprojmatrix} and its inverse~\eqref{eq:invtotprojmatrix} obtaining the relations
\begin{align}
	\left( L^{-1} \right)_0 & = L_0^{-1} + L_0^{-1} L_{0,\rho_1} T L_{\rho_1,0} L_0^{-1} \\
	\left( L^{-1} \right)_{0,\rho_1} & = - L_0^{-1} L_{0,\rho_1} T \\
	\left( L^{-1} \right)_{\rho_1,0} & = - T L_{\rho_1,0} L_0^{-1} \\
	\left( L^{-1} \right)_{\rho_1} & \equiv T = \left( L_{\rho_1} - L_{\rho_1,0} L_0^{-1} L_{0,\rho_1} \right)^{-1} \: .
\end{align}
Identifying the relevant interrelations among the sub-matrices of the inverted projection matrix yields the desired terms
\begin{align}
	\left( L^{-1} \right)_0 & = L_0^{-1} - L_0^{-1} L_{0,\rho_1} \left( L^{-1} \right)_{\rho_1,0} \: , 
	\label{eq:Pn-Pnvac-rel1}\\
	\left( L^{-1} \right)_{0,\rho_1} & = - L_0^{-1} L_{0,\rho_1} \left( L^{-1} \right)_{\rho_1} \: .
	\label{eq:Pn-Pnvac-rel2}
\end{align}
Substituting eqn.~\eqref{eq:Pn-Pnvac-rel1} and \eqref{eq:Pn-Pnvac-rel2} in Eq.~\eqref{eq:Pn-Pnvac} yields
\begin{align}
	\vec l^{(1)}_\mu \cdot \vec a_1
	= \vec l^{(0)}_\mu \cdot L_0^{-1} L_{0,\rho_1} \left( \left( L^{-1} \right)_{\rho_1,0} , \left( L^{-1} \right)_{\rho_1)} \right) \vec l^{(\rho)}_\nu \langle O^\nu \rangle + \vec l^{(\rho_1)}_\mu \cdot \vec a_{\rho_1} \: .
\end{align}
Using Eq.~\eqref{eq:proja(rho)decomp} we arrive at
\begin{align}
	\vec l^{(1)}_\mu \cdot \vec a_1 = -\; \vec l^{(0)}_\mu \cdot L_0^{-1} L_{0,\rho_1} \vec a_{\rho_1} + \vec l^{(\rho_1)}_\mu \cdot \vec a_{\rho_1} \: .
\end{align}
Providing the first term in matrix notation
\begin{align}
	\vec l^{(1)}_\mu \cdot \vec a_1
	= - \left( \vec l^{(0)}_\mu \right)^\mathrm{T} L_0^{-1} L_{0,\rho_1} \vec a_{\rho_1} + \vec l^{(\rho_1)}_\mu \cdot \vec a_{\rho_1}
	= - \left( L_{\rho_1,0}  L_0^{-1} \vec l^{(0)}_\mu \right)^\mathrm{T} \vec a_{\rho_1} + \vec l^{(\rho_1)}_\mu \cdot \vec a_{\rho_1}
\end{align}
finally yields
\begin{align}
	\vec l^{(1)}_\mu \cdot \vec a_1
	= \left[ L_{\rho_1,0} L_0^{-1} \vec l^{(0)}_\mu - \vec l^{(\rho_1)}_\mu\right] \cdot \left( -\; \vec a_{\rho_1} \right)
\end{align}
which is exactly the medium-specific part of the decomposition as obtained from orthogonality of vacuum and medium-specific parts of the decomposition (cf.\ Eq.~\eqref{eq:medprojvec}).
The orthogonality \eqref{eq:defmed2} can be easily shown:
\begin{align}
	\vec l^{(1)} \circ \vec l^{(0)}
	= \left( L_{\rho_1,0} L_0^{-1} \vec l^{(0)} - \vec l^{(\rho_1)} \right) \circ \vec l^{(0)} 
	= L_{\rho_1,0} L_0^{-1} L_0 - L_{\rho_1,0}
	= 0
\label{eq:prooforhto} \: ,
\end{align}
where the orthogonality $\vec l^{(0)} \circ \vec l^{(1)}=0$ can be deduced for the transposed of \eqref{eq:prooforhto}, due to the handy notation of the dyadic product.

\section{Further in-medium decompositions of dimension 6 quark operators}
\label{app:furtherdecomp}

In this section we gather the decompositions of relevant condensates into vacuum and medium-specific parts not explicitly listed in Sec.~\ref{sec:4q}, i.\,e.\ the results for even and odd \gls{OPE} terms \ref{enum:e2}--\ref{enum:e5} and \ref{enum:o2}--\ref{enum:o5} are presented.

\subsection{Even OPE}
\label{app:furtherdecompeven}

\begin{enumerate}[label={\itshape (i--\arabic{*})}, ref={\itshape (i--\arabic{*})}, itemindent=2.5em, leftmargin=0.0em, start=1]
	\label{enum:even2}
	\item The second \gls{EV} \ref{enum:e2} decomposes into vacuum and medium-specific terms as
\begin{align}
	\langle\!\langle \bar{q} \gamma_\mu [D_\nu, G_{\kappa\lambda}] q \rangle\!\rangle_0 & = \frac{1}{12} \left( g_{\mu\kappa}g_{\nu\lambda} - g_{\mu\lambda}g_{\nu\kappa} \right) \langle\!\langle \bar{q} \gamma^{\alpha} [D^{\beta}, G_{\alpha\beta}] q \rangle\!\rangle \: , 
	\label{eq:VacSpecCond2}\\
	\langle\!\langle\bar{q} \gamma_\mu [D_\nu, G_{\kappa\lambda}] q \rangle\!\rangle_1 & = \frac{1}{12} l^{(1)}_{[\mu\nu][\kappa\lambda]} \langle\!\langle \bar{q} \gamma^{\alpha} [D^{\beta}, G_{\alpha\beta}] q \nonumber\\
	&\qquad\qquad\quad- \frac{2}{v^2} \left( \bar{q} \gamma^{\alpha} [(vD), G_{\alpha\beta}] v^\beta q + \bar{q} \slashed{v} [D^{\alpha}, G_{\beta\alpha}] v^\beta q \right) \rangle\!\rangle \: ,
	\label{eq:MedSpecCond2}
\end{align}
where only the part anti-symmetric in both index pairs $\mu\nu$ and $\kappa\lambda$ contributes to the (axial-)vector \gls{OPE}.
Transformation to canonical condensates yields
\begin{align}\label{eq:enum2CanonCondTrafo}
	\left(\begin{array}{l}
				\langle \bar{q} \gamma^{\alpha} [D^{\beta}, G_{\alpha\beta}] q \rangle\\
				\langle \bar{q} \gamma^{\alpha} [(vD), G_{\alpha\beta}] v^\beta q \rangle\\
				\langle \bar{q} \slashed{v} [D^{\alpha}, G_{\beta\alpha}] v^\beta q \rangle
				\end{array}\right)
	=
	\left(\begin{array}{ccc}
				1&0&0\\
				0&1&0\\
				0&0&1
				\end{array}\right)
	\left(\begin{array}{l}
				g \langle \bar{q} \gamma^\mu t^A q \sum_f \bar f \gamma_\mu t^A f \rangle\\
				\langle \bar{q} \gamma^{\alpha} [(vD), G_{\alpha\beta}] v^\beta q \rangle\\
				g \langle \bar{q} \slashed{v} t^A q \sum_f \bar f \slashed{v} t^A f \rangle
				\end{array}\right) \: .
\end{align}

	\label{enum:even3}
	\item The third \gls{EV} \ref{enum:e3} decomposes into vacuum and medium-specific terms as
\begin{align}
	& \langle\!\langle \bar{q} \gamma_5 \gamma_\alpha [D_\sigma, G_{\nu\lambda}] q \rangle\!\rangle_0  = -\frac{1}{24} \varepsilon_{\alpha\sigma\nu\lambda} \langle\!\langle \varepsilon_{\alpha'\sigma'\nu'\lambda'} \bar{q} \gamma_5 \gamma^{\alpha'} [D^{\sigma'}, G^{\nu'\lambda'}] q \rangle\!\rangle \: , \\
	&\langle\!\langle \bar{q} \gamma_5 \gamma_\alpha [D_\sigma, G_{\nu\lambda}] q \rangle\!\rangle_1  \nonumber\\
	& = -\frac{1}{192}
	\left(\begin{array}{l}
				\varepsilon_{\alpha\sigma\nu\lambda} - \frac{4}{v^2} \varepsilon_{\alpha\nu\lambda\tau} v^\tau v_\sigma\\
				- \varepsilon_{\alpha\sigma\nu\lambda} - \frac{4}{v^2} \varepsilon_{\alpha\sigma\lambda\tau} v^\tau v_\nu - ( \varepsilon_{\alpha\sigma\nu\lambda} - \frac{4}{v^2} \varepsilon_{\alpha\sigma\nu\tau} v^\tau v_\lambda )
				\end{array}\right)^\mathrm{T}
	\left(\begin{array}{cc}
				4&-2\\
				-2&3
				\end{array}\right)\nonumber\\
	&\quad\times
	\left(\begin{array}{l}
				\langle\!\langle \varepsilon_{\alpha'\sigma'\nu'\lambda'} \bar{q} \gamma_5 \gamma^{\alpha'} [D^{\sigma'}, G^{\nu'\lambda'}] q - \frac{4}{v^2} \varepsilon_{\alpha'\nu'\lambda'\tau'} v_{\tau'} \bar{q} \gamma_5 \gamma^{\alpha'} [(vD), G^{\nu'\lambda'}] q \rangle\!\rangle\\
				2\langle\!\langle -  \varepsilon_{\alpha'\sigma'\nu'\lambda'} \bar{q} \gamma_5 \gamma^{\alpha'} [D^{\sigma'}, G^{\nu'\lambda'}] q - \frac{4}{v^2} \varepsilon_{\alpha'\sigma'\lambda'\tau'} v_{\tau'} v^{\nu'} \bar{q} \gamma_5 \gamma^{\alpha'} [D^{\sigma'}, G^{\nu'\lambda'}] q \rangle\!\rangle
				\end{array}\right) \: ,
	\label{eq:MedSpecCond3}
\end{align}
where only the part anti-symmetric in the index pair $\nu\lambda$ contributes to the (axial-)vector \gls{OPE}.
Transformation to canonical condensates yields
\begin{align}
	\left(\!\!\begin{array}{l}
					\langle \varepsilon_{\alpha\sigma\nu\lambda} \bar{q} \gamma_5 \gamma^\alpha [D^\sigma, G^{\nu\lambda}] q \rangle\\
					\langle \varepsilon_{\alpha\nu\lambda\tau} v^\tau \bar{q} \gamma_5 \gamma^\alpha [(vD), G^{\nu\lambda}] q \rangle\\
					\langle \varepsilon_{\alpha\sigma\lambda\tau} v^\tau v_\nu \bar{q} \gamma_5 \gamma^\alpha [D^\sigma, G^{\nu\lambda}] q \rangle
				\end{array}\!\!\right)
	=
	\left(\begin{array}{cccc}
				1 &  0 &  0 & 0\\
				0 & -1 & 2i & 0\\
				0 &  0 & -i & i
				\end{array}\right)
	\left(\!\!\begin{array}{l}
					0 \\
					\langle \bar{q} \slashed{v} \sigma^{\nu\lambda} [(vD), G_{\nu\lambda}] q \rangle\\
					\langle \bar{q} \gamma^\lambda [(vD), G_{\nu\lambda}] v^\nu q \rangle\\
					g \langle \bar{q} \slashed{v} t^A q \sum_f \bar f \slashed{v} t^A f \rangle
				\end{array}\!\!\right) \: .
\end{align}

	\label{enum:even4}
	\item The decomposition of \ref{enum:e4} does vanish, when contracted with the Dirac-trace result of the \gls{OPE}.

	\label{enum:even5}
	\item The fifth \gls{EV} \ref{enum:e5} decomposes into vacuum and medium-specific terms as
\begin{align}
	& \langle\!\langle \bar{q} \overset{{}_\leftarrow}{D}_\mu \gamma_5 \gamma_\alpha G_{\nu\lambda} q \rangle\!\rangle_0 = -\frac{1}{24} \varepsilon_{\mu\alpha\nu\lambda} \varepsilon_{\mu'\alpha'\nu'\lambda'} \langle\!\langle \bar{q} \overset{{}_\leftarrow}{D}{}^{\mu'} \gamma_5 \gamma^{\alpha'} G^{\nu'\lambda'} q \rangle\!\rangle \: , 
\end{align}
\begin{align}
	& \langle\!\langle \bar{q} \overset{{}_\leftarrow}{D}_\mu \gamma_5 \gamma_\alpha G_{\nu\lambda} q \rangle\!\rangle_1 \nonumber\\
	& = -\frac{1}{192}
	\left(\begin{array}{l}
				\varepsilon_{\mu\alpha\nu\lambda} - \frac{4}{v^2} \varepsilon_{\mu\nu\lambda\tau} v^\tau v_\alpha\\
				- \varepsilon_{\mu\alpha\nu\lambda} - \frac{4}{v^2} \varepsilon_{\mu\alpha\lambda\tau} v^\tau v_\nu- (\varepsilon_{\mu\alpha\nu\lambda} - \frac{4}{v^2} \varepsilon_{\mu\alpha\nu\tau} v^\tau v_\lambda )
				\end{array}\right)^\mathrm{T}
	\left(\begin{array}{cc}
					 4 & -2 \\
					-2 &  3 \\
				\end{array}\right)\nonumber\\
	& \quad\times
	\left(\begin{array}{l}
				\langle\!\langle \varepsilon_{\mu'\alpha'\nu'\lambda'} \bar{q} \overset{{}_\leftarrow}{D}{}^{\mu'} \gamma_5 \gamma^{\alpha'} G^{\nu'\lambda'} q - \frac{4}{v^2} \varepsilon_{\mu'\nu'\lambda'\tau'} v^{\tau'}\bar{q} \overset{{}_\leftarrow}{D}{}^{\mu'} \gamma_5 \slashed{v} G^{\nu'\lambda'} q \rangle\!\rangle\\
				2 \langle\!\langle - \varepsilon_{\mu'\alpha'\nu'\lambda'} \bar{q} \overset{{}_\leftarrow}{D}{}^{\mu'} \gamma_5 \gamma^{\alpha'} G^{\nu'\lambda'} q - \frac{4}{v^2} \varepsilon_{\mu'\alpha'\lambda'\tau'} v^{\tau'} v_{\nu'} \bar{q} \overset{{}_\leftarrow}{D}{}^{\mu'} \gamma_5 \gamma^{\alpha'} G^{\nu'\lambda'} q \rangle\!\rangle
				\end{array}\right) \: ,
	\label{eq:MedSpecCond4}
\end{align}
where only the part anti-symmetric in the index pair $\nu\lambda$ contributes to the (axial-)vector \gls{OPE}.
Transformation to canonical condensates yields:
\begin{align}
	&\left(\begin{array}{l}
					\langle \varepsilon_{\mu\alpha\nu\lambda} \bar{q} \overset{{}_\leftarrow}{D}{}^{\mu} \gamma_5 \gamma^{\alpha} G^{\nu\lambda} q \rangle\\
					\langle \varepsilon_{\mu\nu\lambda\tau} v^{\tau}\bar{q} \overset{{}_\leftarrow}{D}{}^{\mu} \gamma_5 \slashed{v} G^{\nu\lambda} q \rangle\\
					\langle \varepsilon_{\mu\alpha\lambda\tau} v^{\tau} v_{\nu} \bar{q} \overset{{}_\leftarrow}{D}{}^{\mu} \gamma_5 \gamma^{\alpha} G^{\nu\lambda} q \rangle
				\end{array}\right)
	=
	\left(\begin{array}{ccc}
				i & 0 & 0\\
				0 & 1 & 0\\
				0 & 0 & i
				\end{array}\right)\nonumber\\
	& \qquad\quad \times
	\left(\begin{array}{l}
				m_q \langle \bar{q} \sigma G q \rangle + g \langle \bar{q} \gamma^\mu t^A q \sum_f \bar f \gamma_\mu t^A f \rangle\\
				 \langle \varepsilon_{\mu\nu\lambda\tau} v^{\tau}\bar{q} \overset{{}_\leftarrow}{D}{}^{\mu} \gamma_5 \slashed{v} G^{\nu\lambda} q \rangle\\
				m_q \langle \bar{q} v_\mu \sigma^{\mu\nu} G_{\nu\lambda} v^\lambda q \rangle - \frac{1}{2} g \langle \bar{q} \slashed{v} t^A q \sum_f \bar f \slashed{v} t^A f \rangle - \langle \bar{q} (v\overset{{}_\leftarrow}{D}) \gamma^\nu G_{\nu\lambda} v^\lambda q \rangle
				\end{array}\right) \: .
\end{align}
\end{enumerate}

\subsection{Odd OPE}
\label{app:furtherdecompodd}

\begin{enumerate}[label={\itshape (i--\arabic{*})}, ref={\itshape (i--\arabic{*})}, itemindent=2.5em, leftmargin=0.0em, start=1]
	\label{enum:odd2}
	\item The second \gls{EV} \ref{enum:o2} decomposes as
\begin{align}
	\langle\!\langle \bar q [D_\sigma , G_{\nu\lambda}] q \rangle\!\rangle = \frac{1}{3v^2} (v_\nu g_{\sigma\lambda} - v_\lambda g_{\sigma\nu}) \langle\!\langle \bar q [D^{\mu'} , G_{\nu'\mu'}] v^{\nu'} q \rangle\!\rangle \: ,
\end{align}
where only the part anti-symmetric in $\nu\lambda$ enters the (axial-)vector \gls{OPE}.
Transformation to a canonical condensate yields
\begin{align}
	\langle \bar q [D^{\mu'} , G_{\nu'\mu'}] v^{\nu'} q \rangle = g \langle \bar q t^A q \sum_f \bar f \slashed{v} t^A f \rangle \: .
	\label{eq:chioddcond}
\end{align}

	\label{enum:odd3}
	\item The third \gls{EV} \ref{enum:o3} decomposes as
\begin{align}
	&\langle\!\langle \bar{q} \sigma_{\alpha\beta} [D_\sigma , G_{\nu\lambda}] q \rangle\!\rangle \nonumber\\
	& = 
	\frac{1}{6v^2}
	\left(\begin{array}{l}
					v_\alpha g_{\beta\nu} g_{\sigma\lambda} - v_\alpha g_{\beta\lambda} g_{\sigma\nu} - v_\beta g_{\alpha\nu} g_{\sigma\lambda} + v_\beta g_{\alpha\lambda} g_{\sigma\nu} \\
					v_\sigma ( g_{\alpha\nu} g_{\beta\lambda} - g_{\alpha\lambda} g_{\beta\nu} ) \\
					v_\nu g_{\alpha\sigma} g_{\beta\lambda} - v_\lambda g_{\alpha\sigma} g_{\beta\nu} + v_\lambda g_{\alpha\nu} g_{\beta\sigma} - v_\nu g_{\alpha\lambda} g_{\beta\sigma} \\
					\frac{1}{v^2} ( v_\alpha v_\sigma v_\nu g_{\beta\lambda} - v_\alpha v_\sigma v_\lambda g_{\beta\nu} - v_\beta v_\sigma v_\nu g_{\alpha\lambda} + v_\beta v_\sigma v_\lambda g_{\alpha\nu} )
				\end{array}\right)^\mathrm{T} \nonumber\\
	& \quad \times
	\left(\begin{array}{cccc}
					 1 &  0 &  0 & -1\\
					 0 &  2 &  0 & -2\\
					 0 &  0 &  1 & -1\\
					-1 & -2 & -1 &  6
				\end{array}\right)
	\left(\begin{array}{l}
					\langle\!\langle \bar{q} v_{\alpha'} \sigma^{\alpha'\nu'} [D^{\lambda'} , G_{\nu'\lambda'}] q \rangle\!\rangle\\
					\frac{1}{2} \langle\!\langle \bar{q} \sigma^{\nu'\lambda'} [(vD) , G_{\nu'\lambda'}] q \rangle\!\rangle\\
					\langle\!\langle \bar{q} \sigma^{\sigma'\lambda'} [D_{\sigma'} , G_{\nu'\lambda'}] v^{\nu'} q \rangle\!\rangle\\
					\frac{1}{v^2} \langle\!\langle \bar{q} v_{\alpha'} \sigma^{\alpha'\lambda'} [(vD) , G_{\nu'\lambda'}] v^{\nu'} q \rangle\!\rangle
				\end{array}\right) \: ,
	\label{eq:MedSpecCondodd3}
\end{align}
where only the combination $\langle\!\langle \bar{q} \sigma^{\nu'\lambda'} [(vD) , G_{\nu'\lambda'}] q + \bar{q} \sigma^{\sigma'\lambda'} [D_{\sigma'} , G_{\nu'\lambda'}] v^{\nu'} q \rangle\!\rangle$ enters the (axial-)vector \gls{OPE}.
Application of the gluon equation of motion to the first condensate in \eqref{eq:MedSpecCondodd3} yields the four-quark condensate $\langle\!\langle \bar{q} v_{\alpha'} \sigma^{\alpha'\nu'} t^A q \sum_f \bar f \gamma_{\nu'} t^A f \rangle\!\rangle$ which is not invariant under parity and time reversal transformations. The other condensates in \eqref{eq:MedSpecCondodd3} cannot be reduced by equations of motion.

	\label{enum:odd4}
	\item The forth \gls{EV} \ref{enum:o4} decomposes as
\begin{align}
	\langle\!\langle \bar q \overset{{}_\leftarrow}{D}_\mu G_{\nu\lambda} q \rangle\!\rangle = \frac{1}{3v^2} (v_\nu g_{\mu\lambda} - v_\lambda g_{\mu\nu}) \langle\!\langle \bar q \overset{{}_\leftarrow}{D}{}^{\mu'}  G_{\nu'\mu'} v^{\nu'} q \rangle\!\rangle \: ,
\end{align}
where only the part anti-symmetric in $\nu\lambda$ enters the (axial-)vector \gls{OPE}.
Transformation to a canonical condensate yields
\begin{align}
	\langle \bar q \overset{{}_\leftarrow}{D}{}^{\mu'}  G_{\nu'\mu'} v^{\nu'} q \rangle = \frac{g}{2} \langle \bar q t^A q \sum_f \bar f \slashed{v} t^A f \rangle \: .
\end{align}

	\label{enum:odd5}
	\item The fifth \gls{EV} \ref{enum:o5} decomposes as
\begin{align}
	&\langle\!\langle \bar{q} \overset{{}_\leftarrow}{D}_\mu \sigma_{\alpha\beta} G_{\nu\lambda} q \rangle\!\rangle \nonumber\\
	& = 
	\frac{1}{6v^2}
	\left(\begin{array}{l}
					v_\alpha g_{\beta\nu} g_{\mu\lambda} - v_\alpha g_{\beta\lambda} g_{\mu\nu} - v_\beta g_{\alpha\nu} g_{\mu\lambda} + v_\beta g_{\alpha\lambda} g_{\mu\nu} \\
					v_\mu ( g_{\alpha\nu} g_{\beta\lambda} - g_{\alpha\lambda} g_{\beta\nu} ) \\
					v_\nu g_{\alpha\mu} g_{\beta\lambda} - v_\lambda g_{\alpha\mu} g_{\beta\nu} + v_\lambda g_{\alpha\nu} g_{\beta\mu} - v_\nu g_{\alpha\lambda} g_{\beta\mu} \\
					\frac{1}{v^2} ( v_\alpha v_\mu v_\nu g_{\beta\lambda} - v_\alpha v_\mu v_\lambda g_{\beta\nu} - v_\beta v_\mu v_\nu g_{\alpha\lambda} + v_\beta v_\mu v_\lambda g_{\alpha\nu} )
				\end{array}\right)^\mathrm{T} \nonumber\\
	& \quad \times
	\left(\begin{array}{cccc}
					 1 &  0 &  0 & -1\\
					 0 &  2 &  0 & -2\\
					 0 &  0 &  1 & -1\\
					-1 & -2 & -1 &  6
				\end{array}\right)
	\left(\begin{array}{l}
					\langle\!\langle \bar{q} \overset{{}_\leftarrow}{D}{}^{\lambda'}v_{\alpha'} \sigma^{\alpha'\nu'} G_{\nu'\lambda'} q \rangle\!\rangle\\
					\frac{1}{2} \langle\!\langle \bar{q} (v\overset{{}_\leftarrow}{D}) \sigma G q \rangle\!\rangle\\
					\langle\!\langle \bar{q} \overset{{}_\leftarrow}{D}_{\alpha'} \sigma^{\sigma'\lambda'} G_{\nu'\lambda'} v^{\nu'} q \rangle\!\rangle\\
					\frac{1}{v^2} \langle\!\langle \bar{q} (v\overset{{}_\leftarrow}{D}) v_{\alpha'} \sigma^{\alpha'\lambda'} G_{\nu'\lambda'} v^{\nu'} q \rangle\!\rangle
				\end{array}\right) \: ,
	\label{eq:MedSpecCondodd5}
\end{align}
where only the term containing $\langle\!\langle \bar{q} (v\overset{{}_\leftarrow}{D}) \sigma G q \rangle\!\rangle$ enters the (pseudo-)scalar \gls{OPE}.
Application of gluon equation of motion to the first condensate in \eqref{eq:MedSpecCondodd5} yields $\langle\!\langle \bar{q} v_{\alpha'} \sigma^{\alpha'\nu'} t^A q \sum_f \bar f \gamma_{\nu'} t^A f \rangle\!\rangle$ which is not invariant under parity and time reversal transformations.
The other condensates in \eqref{eq:MedSpecCondodd5} cannot be reduced by equations of motion.
\end{enumerate}

\bibliographystyle{aip}
\bibliography{lit}

\end{document}